\documentclass[aps,superscriptaddress,floatfix,floats,superscriptaddress]{revtex4}

\usepackage{amsmath}
\usepackage{amssymb}
\usepackage{amsthm}
\usepackage{amsfonts}

\usepackage{graphicx}
\usepackage[colorlinks,bookmarks=false]{hyperref}
\usepackage{bbm}

\newcommand{\id}[1] {\ensuremath{\text{d} #1 \;}}
\newcommand{\be}{\begin{equation}}
\newcommand{\ee}{\end{equation}}
\newcommand{\bea}{\begin{eqnarray}}
\newcommand{\eea}{\end{eqnarray}}
\def\la{\lambda}
\def\nn{\nonumber\\}
\newcommand{\ket}[1]{\ensuremath{\left| #1 \right>}}

\providecommand{\Tr}[1]{\ensuremath{\text{Tr}\hspace*{-2pt} \left(#1\right) }}
\newcommand{\tr}[1]{\ensuremath{\text{Tr}\left( #1 \right) }}
\renewcommand{\exp}[1]{\ensuremath{ \; \text{exp} \left( #1 \right) } }
\newcommand{\abs}[1]{\left| #1 \right|}
\newcommand{\avg}[1]{\left< #1 \right>}
\renewcommand{\det}[1]{\ensuremath{\text{det}\left( #1 \right) }}
\newcommand{\eps}{\ensuremath{\varepsilon}}
\def\fr#1{(\ref{#1})}

\begin{document}
%%%%%%%%%%%%%%%%%%%%%%%%%%%%%%%%%%%%%%%%%%%%%%%%%%%%%%%%%%%%%%%%%%%%%
\title{Full Counting Statistics in the Transverse Field Ising Chain}
%%%%%%%%%%%%%%%%%%%%%%%%%%%%%%%%%%%%%%%%%%%%%%%%%%%%%%%%%%%%%%%%%%%%%
\author{Stefan \surname{Groha}}
\author{Fabian \surname{Essler}} 
\affiliation{The Rudolf Peierls Centre for Theoretical Physics, Oxford University, Oxford, OX1 3NP, UK}
\author{Pasquale \surname{Calabrese}}
\affiliation{SISSA and INFN, via Bonomea 265, 34136 Trieste, Italy}
\affiliation{International Centre for Theoretical Physics (ICTP), I-34151, Trieste, Italy}
\begin{abstract}
We consider the full probability distribution for the transverse
magnetization of a finite subsystem in the transverse field Ising chain.
We derive a determinant representation of the corresponding
characteristic function for general Gaussian states. We consider
applications to the full counting statistics in the ground state,
finite temperature equilibrium states, non-equilibrium steady states
and time evolution after global quantum quenches. We derive an
analytical expression for the time and subsystem size dependence of
the characteristic function at sufficiently late times after a quantum
quench. This expression features an interesting multiple light-cone structure.
\end{abstract}
\date{\today}
\maketitle

%%%%%%%%%%%%%%%%%%%%%%%%%%%%%%%%%%%%%%%%%%%%%%%%%%%%%%%%%%%%%%%%%%%%%%%%%%
%%%%%%%%%%%%%%%%%%%%%%%%%%%%%%%%%%%%%%%%%%%%%%%%%%%%%%%%%%%%%%%%%%%%%%%%%%
\section{Introduction}
%%%%%%%%%%%%%%%%%%%%%%%%%%%%%%%%%%%%%%%%%%%%%%%%%%%%%%%%%%%%%%%%%%%%%
The statistical nature of measurements of observables is a fundamental
principle of quantum mechanics. Measuring the same observable in
identically prepared systems leads to different measurement outcomes
that are described by a probability distribution that depends on both the
state $|\Psi\rangle$ and on the observable ${\cal O}$
considered. The full probability distribution $P({\cal
  O},|\Psi\rangle)$ encodes detailed information about quantum
fluctuations in the system. It is of particular interest in situations
where the first few moments do not provide a good description of the
distribution. Quantum mechanical probability distributions in the
guise of \emph{Full Counting Statistics} (FCS) have been
studied for some time in mesoscopic devices\cite{CM,CM2}. More
recently it has become possible to analyze them in systems of
ultra-cold atomic gases
\cite{HLSI08,KPIS10,KISD11,GKLK12,AJKB10,JABK11}. This has broken new
ground in the sense that one is dealing with (strongly) interacting
many-particle systems and a variety of observables, typically defined
on subsystems, can be accessed. This has motivated a number of
theoretical works of FCS in equilibrium states
\cite{cd-07,lp-08,ia-13,sk-13,e-13,k-14,mcsc-15,sp-17,CoEG17,nr-17,hb-17,bpc-18},
and after quantum quenches \cite{gadp-06,IGD,er-13,KISD11,lddz-15,bpc-18}.
A second motivation for studying FCS has been the observation that in
non-interacting fermionic systems with particle number conservation
the FCS of particle number within a subsystem is directly related to
the entanglement entropy
\cite{kl-09a,kl-09b,hgf-09,sfr-11a,sfr-11b,cmv-12,lbb-12,si-13,clm-15,hml-17} and provides
indirect information about the latter.

From a theoretical point of view calculating the FCS for a given
observable on a sizeable subsystem poses a formidable problem and
as a result only very few exact results are available even in simple
equilibrium situations. Even less is known about FCS after quantum
quenches. This motivates reconsidering FCS in the transverse field
Ising chain (TFIC). The TFIC is a key paradigm for quantum
phase transitions \cite{sach-book} and a simple, but non-trivial, many-body system without
particle number conservation and therefore provides an ideal
playground for studying FCS both in and out of equilibrium. Indeed,
thanks to the mapping of the TFIC to a model of non-interacting
spinless fermions with pairing term it is possible to analytically
determine ground state and thermal properties, see
e.g. \cite{BM71a,BM71b,sach-book}, as well as describe the
non-equilibrium dynamics of local observables
\cite{mc,ir-00,sps-04,CEF1,CEF2,CEF3,EEF:12} and of the reduced density
matrix of a block of adjacent sites \cite{cc-05,FC08,CEF3,FE_13a} 
after a global quantum quench. A summary of these developments is
given in the recent reviews Refs~\cite{cem-16,EFreview}. 

In this work we focus on the FCS of the simplest observable, the
transverse magnetization within a block of $\ell$ adjacent spins.  
In the ground state this problem has been
previously analysed in Refs~\cite{cd-07,ia-13} and generic Gaussian
states have been considered as well \cite{k-14}. We note that the
ground state FCS of the longitudinal magnetization at the critical
point has been determined in Ref.~\cite{lp-08} and the ground state
FCS of the subsystem energy was considered in Ref.~\cite{nr-17}.

This manuscript is organised as follows. 
In section \ref{sec:FCS} we first introduce the TFIC
and briefly summarize the important steps for diagonalizing the
Hamiltonian. We then define the FCS and the associated generating
functions considered in this work. In section \ref{Sec2} we provide a
novel derivation of an efficient determinant representation for the
FCS in general $\mathbb{Z}_2$ invariant Gaussian states. The 
result is equivalent to that of Ref.~\cite{k-14}. This result is
applied in section \ref{Sec3} to the determination of the FCS in
equilibrium states. In the ground state we recover the results of
Ref.~\cite{cd-07}. Our results for the FCS in finite temperature
equilibrium states are to the best of our knowledge new. In section
\ref{sec:QQ} we turn to the main point of interest: the time evolution
of the FCS after a global quantum quench. We consider the situation
where the system is prepared in a pure state at a finite finite energy
density and then time evolved with a Hamiltonian $H$ that does not
commute with the initial state density matrix, which leads to
non-trivial dynamics. We present explicit results for general
``transverse field'' quenches as well as evolution starting in a
classical N\'eel state. The main result of this work is presented in
section \ref{sec:analytic}: an analytic expression for the time
evolution of the FCS after a transverse field quench.
In section \ref{sec:summ} we summarize our results and comment on a
number of issues that deserve further investigation.

\section{The model and the full counting statistics}
\label{sec:FCS}

%%%%%%%%%%%%%%%%%%%%%%%%%%%%%%%%%%%%%%%%%%%%%%%%%%%%%%%%%%%%
\subsection{Transverse Field Ising chain}
\label{ssec:TFIC}
%%%%%%%%%%%%%%%%%%%%%%%%%%%%%%%%%%%%%%%%%%%%%%%%%%%%%%%%%%%%
In the following we consider the spin-$1/2$ transverse field Ising
model on an infinite chain
\begin{align}
H(h) =& 
-\sum_{j=-\infty}^\infty \left[ \sigma_j^x \sigma_{j+1}^x +h \sigma_j^z\right].
 \label{His} 
\end{align}
The ground state phase diagram features ferromagnetic ($h<1$) and
paramagnetic ($h>1$) phases that are separated by a quantum critical
point in the universality class of the two-dimensional Ising model
\cite{sach-book}. The order parameter that characterizes the
transition is the longitudinal magnetisation $\langle{\rm
  GS}|\sigma^x_j|{\rm GS}\rangle$. At finite temperature spontaneous
breaking of the $\mathbbm{Z}_2$ symmetry of $H(h)$ is forbidden and
hence the order present in the ground state at $h<1$ melts.
In order for this paper to be self-contained we now briefly summarize
the relevant steps for diagonalizing the Hamiltonian \fr{His}. A more
detailed discussion can be found in e.g. the Appendix in
\cite{CEF2}. The TFIC is mapped to a model of spinless fermions by a
Jordan-Wigner transformation
\be
\sigma_j^z= 1-2c^\dagger_j c^{\phantom{\dagger}}_j\ ,\qquad
\sigma_j^x=\prod_{l=-\infty}^{j-1} (1-2c_l^\dagger
c_l^{\phantom{\dagger}})
(c_j+c_j^\dagger)\ ,
\label{JW}
\ee
where $c_j$ are fermion operators obeying canonical anticommutation
relations $\{c^\dagger_j,c_k\}=\delta_{j,k}$. Setting aside
the issue of boundary conditions the Hamiltonian takes the form
\begin{align}
  H(h) =& -J \sum_{j=-\infty}^{\infty} (c_j^\dagger - c_j)(c_{j+1}+c_{j+1}^\dagger) -Jh( c^{\phantom{\dagger}}_jc^\dagger_j-c_j^\dagger c^{\phantom{\dagger}}_j).
\end{align}
This is diagonalized by a Bogoliubov transformation
\begin{align}
  c_j = \int_{-\pi}^\pi \frac{dk}{2\pi} e^{-ik j} 
\left[\cos(\theta_{k}/2) \alpha_{k} +i \sin(\theta_{k}/2)
  \alpha^\dagger_{-k}\right],
\end{align}
where $\{\alpha_k,\alpha^\dagger_p\}=\delta_{p,k}$ and
the Bogoliuobov angle is 
\begin{align}
  e^{i\theta_k} = \frac{h-e^{ik}}{\sqrt{1+h^2-2h\cos k}}\ .
\label{Bogangle}
\end{align}
The Hamiltonian takes the form
\begin{align}
  H(h) =& \int_{-\pi}^\pi \frac{dk}{2\pi} \eps(k) \left[ \alpha^\dagger_k \alpha_k -\frac{1}{2}\right],
\end{align}
where the dispersion relation is given by
\begin{align}
\eps(k) =& 2J\sqrt{1+h^2-2h\cos(k)}.
\label{dispersion}
\end{align}
The ground state of $H(h)$ is equal to the Bogoliubov vacuum state
defined by
\begin{align}
  \alpha_k \ket{0} = 0.
\end{align}

%%%%%%%%%%%%%%%%%%%%%%%%%%%%%%%%%%%%%%%%%%%%%%%%%%%%%%%%%%%%%%
\subsection{Full Counting Statistics and Generating Function} 
%%%%%%%%%%%%%%%%%%%%%%%%%%%%%%%%%%%%%%%%%%%%%%%%%%%%%%%%%%%%%%  
We are interested in the properties of the smooth and staggered
components of the transverse magnetization of a chain segment of
length $\ell$. These are defined as
\be
S_u^z(\ell)=\sum_{j=1}^\ell\sigma^z_j\ ,\qquad
S_s^z(\ell)=\sum_{j=1}^\ell(-1)^j\sigma^z_j\ .
\ee
Given a density matrix $\rho$ that specifies the quantum mechanical
state of our system, the probability distributions for the transverse
subsystem magnetizations are given by
\begin{align}
P^{({\rm u,s})}(m) = \tr{\rho\, \delta\big(m-S_{u,s}^z(\ell)\big)}\ .
\label{Pus}
\end{align}
In the following we will focus on the characteristic functions of
these probability distributions, defined as
\bea
P^{\rm(u,s)}(m) &=& \int_{-\infty}^\infty \frac{d\lambda}{2\pi}\
 e^{-i\lambda m}\ \chi^{\rm(u,s)}(\lambda,\ell)\ ,\nn
\chi^{\rm (u,s)}(\lambda,\ell) &=& {\rm Tr}\left[\rho\
 e^{i\lambda S^z_{u,s}}\right].
\label{chidef}
\eea
By construction, the expansion of $\chi^{\rm (u,s)}(\lambda,\ell)$
around $\lambda=0$ generates the moments of the associated probability
distribution. The following relations are readily inferred from the
definition of $\chi^{\rm (u,s)}(\lambda,\ell)$
\bea
\chi^{\rm (u,s)}(\lambda,\ell)&=&\left[\chi^{\rm
    (u,s)}(-\lambda,\ell)\right]^*\ ,\nn
\chi^{\rm (u,s)}(0,\ell)&=&1\ ,\nn
\chi^{\rm (u,s)}(\lambda+\pi,\ell)&=& (-1)^\ell 
\chi^{\rm (u,s)}(\lambda,\ell)\ .
\eea
These properties imply
\be
P^{\rm(u,s)}(m)=2\sum_{r\in\mathbb{Z}} P^{\rm(u,s)}_w(r)
\begin{cases} 
\delta(m-2r+\ell) & \text{if }\ell \text{ is odd} \\  
\delta(m-2r)  & \text{if }\ell \text{ is even}\end{cases}
\ee
where we have defined the weights
\begin{align}
P^{\rm(u,s)}_w(r) = \int_{-\pi/2}^{\pi/2} \frac{d\lambda}{2\pi}
e^{-2i\lambda r} \chi^{\rm (u,s)}(\lambda,\ell)\ .
\label{Pweights}
\end{align}

%%%%%%%%%%%%%%%%%%%%%%%%%%%%%%%%%%%%%%%%%%%%%%%%%%%%%%%%%%%
\section{Generating Function for a general gaussian state}
\label{Sec2}
%%%%%%%%%%%%%%%%%%%%%%%%%%%%%%%%%%%%%%%%%%%%%%%%%%%%%%%%%%%
In this section we show how to obtain the generating function 
\eqref{chidef} for a general Gaussian state with a novel method that
is however equivalent to the one used in \cite{k-14}.

Our starting point is the realization that \eqref{chidef} depends only
on the reduced density matrix of the block $A$ of $\ell$ adjacent
spins 
\be
\chi^{(u,s)}(\lambda,\ell) = {\rm Tr}\left[\rho\ 
e^{i\lambda S^z_{u,s}(\ell)}\right]={\rm
  Tr}\left[\rho_A\ e^{i\lambda S^z_{u,s}(\ell)}\right] \equiv
\widetilde{Z}\ {\rm Tr}\left[\rho_A\ \widetilde{\rho}^{(u,s)}\right]\ ,\quad
a=u,s,
  \label{repr}
\ee
where we have introduced the auxiliary ``density matrices''
\be
\widetilde{\rho}^{(u,s)} \equiv \frac{1}{{\widetilde Z}^{(u,s)}} \
e^{i\lambda S^z_{u,s}(\ell)},\qquad
\widetilde{Z}^{(u,s)} ={\rm Tr}\left[e^{i\lambda S^z_{u,s}(\ell)}\right]
=  \left(2\cos(\lambda)\right)^\ell.
\label{fake1}
\ee
Here the ``partition function'' $\widetilde{Z}^{(a)}$ ensures the normalisation
$\tr{\widetilde{\rho}^{(a)}}=1$. A fundamental property that we will exploit in
the following is that both $\rho_A$ and $\widetilde{\rho}^{(a)}$
are Gaussian operators in the fermionic representation of our
problem, \emph{cf.} section \ref{ssec:TFIC}. Hence they are univocally
determined by the correlation matrices of the fundamental
fermionic operators \cite{peschel2003,lrv-03,pe-09}. Moreover, the
trace of the product of Gaussian operators such as \eqref{repr}
can be expressed in terms of the associated correlation matrices
\cite{fc-10}. This is a very useful property, see
e.g. Ref.~\cite{FE_13a} for a related application, that forms the
basis of our analysis.

In order to proceed we need to specify a convenient basis of
operators. This is provided by Majorana fermions related to the
lattice spin operators by
\be
a_{2l-1}= \left( \prod_{m<l} \sigma_m^z \right) 
\sigma_l^x\ , \qquad 
a_{2l}= \left( \prod_{m<l} \sigma_m^z \right) \sigma_l^y\ , 
\qquad\sigma_l^z= ia_{2j}a_{2j-1}.
\ee
The Majorana fermions satisfy the algebra
\be
\{a_j,a_k\} = 2\delta_{j,k}\ .
\ee
They are related to the Jordan-Wigner fermions \eqref{JW} by
$a_{2l-1} = c^\dag_l + c_l$ and  $a_{2l} = -i(c^\dag_l - c_l)$.

As we are dealing with Gaussian density matrices we can follow Refs.
\cite{peschel2003,lrv-03,pe-09} and Wick's theorem to express $\rho_A$
in terms of the subsystem \emph{correlation matrix} $\Gamma^A_{nm}$ 
\begin{align}
\Gamma_{nm}^A= {\rm Tr}\left[\rho\ a_ma_n\right] -\delta_{nm}\ ,\qquad
1\leq m, n \leq 2\ell.
\label{gammaA}
\end{align}
As the Pauli matrices form a basis in the space of operators over
$\mathbb{C}^2$ the reduced density matrix of a subsystem $A$ that consists
of $\ell$ neighbouring spins at sites  $i=1,\dots,\ell$ can be
expressed in the form
\begin{align}
\rho_A = \frac{1}{2^\ell} \sum_{\{\alpha_1\dots\alpha_\ell\}} 
\tr{\rho\ \sigma_{1}^{\alpha_1}\dots \sigma_{\ell}^{\alpha_\ell}} 
\sigma_{1}^{\alpha_1}\dots \sigma_{\ell}^{\alpha_\ell}\ ,
\end{align}
where $\alpha_i=0,x,y,z$. We now restrict our discussion tor density
matrices that are invariant under the $\mathbbm{Z}_2$ transformation
\be
P\sigma_l^zP=\sigma_l^z\ ,\qquad
P\sigma_l^{x,y}P=-\sigma_l^{x,y}\ .
\ee
In this case the Jordan-Wigner strings cancel and the reduced density
matrix (RDM) is mapped to an operator expressed in terms of Majorana
fermions acting on the same spatial domain
\begin{align}
\rho_A = \frac{1}{2^\ell} \sum_{\{\mu_1\dots \mu_\ell=0,1\}}
\tr{\rho\ a_1^{\mu_1}\dots a_{2\ell}^{\mu_{2\ell}}}
a_{2\ell}^{\mu_{2\ell}}\dots a_1^{\mu_1}\ . 
\label{RDMa}
\end{align}
We note that the case where $P\rho P\neq\rho$ can be dealt with by the
method set out in Ref.~\cite{FE_13a}.
The RDM \fr{RDMa} can be written in an explicit Gaussian form as
\be
\rho_A = \frac{1}{Z} {\rm exp}\left[\frac{1}{4}\sum_{m,n} a_m
  W_{mn}a_n\right], 
\ee
where $W$ is a skew symmetric $2\ell \times 2\ell$ hermitian matrix. 
Using Wick's theorem the matrix $W$ can be related to the correlation
matrix \fr{gammaA}
\begin{align}
\tanh\frac{W}{2} = \Gamma^A.
\end{align}
The auxiliary density matrices $\widetilde{\rho}^{(u,s)}$
\eqref{fake1} can be expressed in the Majorana basis in a completely
analogous way. The corresponding $2\ell\times2\ell$ correlation matrices
$\widetilde{\Gamma}^{(u,s)}$ are given by
\bea
\widetilde{\Gamma}^{(u)}_{ij}&=&{\rm Tr}\left[\widetilde\rho^{(u)}\ a_ja_i\right]=
\frac1{\widetilde{Z}^{(u)}}\Tr {\prod_{k=1}^\ell (\cos\la- i\sin\la
  a_{2k}a_{2k-1}) a_j a_i}-\delta_{ij}\ ,\nn
\widetilde{\Gamma}^{(s)}_{ij} &=&{\rm Tr}\left[\widetilde\rho^{(s)}\ a_ja_i\right]=
 \frac{1}{\widetilde{Z}^{(s)}} {\rm Tr}\left[\prod_{k=1}^\ell
\left(\cos\left(\lambda\right) -i(-1)^k\sin(\lambda) \; a_{2k}a_{2k-1}\right)a_j a_i\right]-\delta_{ij}.
\eea
The only non-vanishing matrix elements are
\bea
\widetilde\Gamma^{(u)}_{2j,2j-1}&=&-\widetilde{\Gamma}^{(u)}_{2j-1,2j}
=\frac{1}{2\cos\la} {\rm Tr}\left[(\cos\la- i\sin\la \,a_{2j}a_{2j-1}) a_{2j-1}
  a_{2j}\right]=-\tan\la\ ,\nn
\widetilde\Gamma^{(s)}_{2j,2j-1}&=&-\widetilde{\Gamma}^{(s)}_{2j-1,2j}
=\frac{1}{2\cos\la} {\rm Tr}\left[(\cos\la- i(-1)^j\sin\la
  \,a_{2j}a_{2j-1}) a_{2j-1} a_{2j}\right]=-(-1)^j\tan\la\ .
 \label{tgammaus}
\eea
This implies that $\widetilde{\Gamma}^{(u,s)}$ are block-diagonal, e.g.
\bea
\widetilde{\Gamma}^{(u)} &=&i \tan \la \left[
 \begin{array}{ccc}
 \sigma_y & 0 & \dots\\
 0 & \sigma_y & \dots\\
 && \ddots
 \end{array}
 \right]\equiv i \tan(\la) \Sigma_y,
 \label{tgamma_f}
\eea
where $\sigma_y$ is the $2\times 2$ Pauli matrix. 

We are now in a position to write down a convenient determinant
representation for the generating functions
$\chi^{(u,s)}(\lambda,\ell)$. To do so we employ a relation derived in
Ref.~\cite{fc-10}: given two Gaussian density matrices $\rho_{1,2}$
with correlation matrices $\Gamma_{1,2}$ the trace of
their product is given by
\begin{align}
{\rm Tr}\left[\rho_1\ \rho_2\right] =\sqrt{
\det{\frac{1+\Gamma_1\Gamma_2}{2}}}\ .
\end{align}
Applying this relation to our case we arrive at the following
determinant representations
\be
\chi^{(a)}(\lambda,\ell)= 
\frac1{ (2\cos \lambda)^\ell}
\sqrt{\det{\frac{1+\Gamma^A\widetilde{\Gamma}^{(a)}}{2}}}\ ,\quad
a=u,s\ ,
\label{mainchi}
\ee
where $\Gamma^A$ and $\Gamma^{(u,s)}$ are given in \eqref{gammaA} and
\eqref{tgammaus}, \fr{tgamma_f} respectively.

%%%%%%%%%%%%%%%%%%%%%%%%%%%%%%%%%%%%%%%%%%%%
\subsection{Simplification in special cases}
%%%%%%%%%%%%%%%%%%%%%%%%%%%%%%%%%%%%%%%%%%%%
Equation \eqref{mainchi} has been derived for a general
$\mathbbm{Z}_2$-invariant Gaussian state with density matrix
$\rho$. If the state is also invariant under translations 
and reflections with respect to a site the generating function
$\chi^{(u)}(\lambda,\ell)$ can be simplified further. 
Indeed, under these conditions, the correlation matrix assumes a
block Toeplitz form \cite{lrv-03,cc-05} 
\begin{align}
  \Gamma^A=\begin{pmatrix} \Pi_0 & \Pi_{-1} & \dots & \Pi_{1-\ell} \\
    \Pi_1 & \Pi_0 & & \vdots \\
    \vdots & & \ddots & \vdots \\
    \Pi_{\ell-1} &\dots&\dots& \Pi_0
  \end{pmatrix}, \qquad \Pi_l = \begin{pmatrix} -f_l & g_l \\ -g_{-l} & f_l
  \end{pmatrix},
  \label{Btoe}
\end{align}
where
\bea
g_l &=& {\rm Tr}\big(a_{2n} a_{2n+2l-1}\big) = 
-{\rm Tr}\big(a_{2n-1}a_{2n-2l}\big)\ ,\nn
f_l &=& {\rm Tr}\big( a_{2n}a_{2n+2l}\big)-\delta_{l0}\ .
\eea
Taking advantage of the block diagonal form of the correlation matrix
of the auxiliary density matrix in \eqref{tgamma_f} we can cast
the generating function in the form
\be
\chi^{(u)}(\lambda,\ell)= (2\cos\la)^\ell \sqrt{ 
\det{\frac{ 1-\tan(\lambda) \Gamma'}{2}}}\ ,
\label{chiTI}
\ee
where $\Gamma^\prime$ is a block Toeplitz matrix 
\begin{align}
\Gamma'=\begin{pmatrix} \Pi'_0 & \Pi'_{-1} & \dots & \Pi'_{1-\ell} \\
    \Pi'_1 & \Pi'_0 & & \vdots \\
    \vdots & & \ddots & \vdots \\
    \Pi'_{\ell-1} &\dots&\dots& \Pi'_0
  \end{pmatrix}, \qquad
\Pi^\prime_l &= \begin{pmatrix} g_l & f_l \\ f_l & g_{-l} \end{pmatrix}.
\label{blockTI}
\end{align}

%%%%%%%%%%%%%%%%%%%%%%%%%%%%%%%%%%%%%%%%%%%%%%%%%%%%%%%%
\subsection{Expressions for the first few cumulants}
%%%%%%%%%%%%%%%%%%%%%%%%%%%%%%%%%%%%%%%%%%%%%%%%%%%%%%%%
The determinant representation \eqref{mainchi} of the generating
function provides an efficient way for determining the cumulants of
the probability distribution, which is the main purpose of the
function itself. The cumulants are obtained in the usual way from the
series expansions of $\ln\chi^{(u,s)}(\lambda,\ell)$
\begin{align}
\ln\chi^{(u,s)}(\lambda,\ell) =& \sum_{n=1}^\infty
\frac{C^{(u,s)}_n}{n!}(i\lambda)^n. 
\end{align}
The first few terms of the series expansion are
\bea
\ln \chi^{(u)}(\lambda,\ell) &=& 
\ell \ln (\cos\lambda) - \frac12 \sum_{n=1}^\infty
\frac{(\tan\lambda)^n}n {\rm Tr}\left[(\bar{\Gamma})^n\right]\nn
&=&-\ell \frac{\lambda^2}2 
-\frac{\lambda}2 \tr{\bar{\Gamma}}-\frac{\lambda^2}4 
\tr{\bar{\Gamma}^2}- \frac{\lambda^3}6 (\tr{\bar{\Gamma}^3} + 
\tr{\bar{\Gamma}}) +O(\lambda^4)\ ,
\eea
where we have defined
\be
\bar\Gamma=-i\Gamma_A\Sigma_y\ .
\ee
The first three cumulants are 
\be
C_1= \frac{i}2 \tr{\bar{\Gamma}}\ ,\quad
C_2= \ell+ \frac{\tr{\bar{\Gamma}^2}}2\ ,\quad
C_3=-\frac{i}2(\tr{\bar{\Gamma}^3} + \tr{\bar{\Gamma}})).
\ee
Specifying to the case of density matrices $\rho$ that are invariant
under translations and reflections around a site we have
\begin{align}
  \tr{\bar{\Gamma}}=&\ell \, \tr{\Pi'_0}= 2\ell g_0, \\
 \tr{\bar{\Gamma}^2}=&\ell \,{\rm Tr}\bigg({\sum_{j=0}^{\ell-1} (2(\ell-j)-\ell\delta_{j0}) \Pi'_j  \Pi'_{-j}} \bigg)
 %\notag \\ =& 
 =2 \ell\sum_{j=0}^{\ell-1}(2(\ell-j)-\ell \delta_{j0}) (g_jg_{-j}+f_j f_{-j}) 
 \notag \\=& 
 2 \ell\bigg(\sum_{j=1}^{\ell-1}(2(\ell-j)) (g_jg_{-j}+f_j f_{-j}) +\ell g_0^2\bigg)
 %\notag \\=& 
 =2 \tr{F^2+G^2}, \\
\tr{\bar{\Gamma}^3}=& 2 \tr{G^3+ 3 F^2G},
\end{align}
where $F$ and $G$ are the $\ell\times\ell$ Toeplitz matrices 
\be
G=\begin{pmatrix} 
g_0 & g_{-1} & \dots & g_{1-\ell} \\
g_1 & g_0 & & \vdots \\
\vdots & & \ddots & \vdots \\
g_{\ell-1} &\dots&\dots& g_0
\end{pmatrix}, \qquad
F=\begin{pmatrix} 
f_0 & f_{-1} & \dots & f_{1-\ell} \\
f_1 & f_0 & & \vdots \\
\vdots & & \ddots & \vdots \\
f_{\ell-1} &\dots&\dots& f_0
\end{pmatrix}\ .
\label{FG}
\ee
For the first three cumulants we obtain  
\be
C_1=i \ell g_0\ ,\quad
C_2= \ell+ \tr{F^2+G^2}\ ,\quad
C_3= - i (\tr{G^3+ 3 F^2G} +\ell g_0)\ .
\ee
It is straightforward to generalise these considerations to higher
cumulants because $\tr{\bar\Gamma^n}$ can always be written as the
sum of the traces of products of $F$ and $G$.  

%%%%%%%%%%%%%%%%%%%%%%%%%%%%%%%%%%%%%%%%%%%%%%%%%%%%
\section{Full counting statistics in equilibrium}
\label{Sec3}
%%%%%%%%%%%%%%%%%%%%%%%%%%%%%%%%%%%%%%%%%%%%%%%%%%%%
In this section we analyze the generating function
$\chi^{(u)}(\lambda,\ell)$ obtained from \eqref{mainchi} and the
associated probability distribution in equilibrium configurations.
We first consider the ground state FCS, which has been previously
studied  by Cherng and Demler in \cite{cd-07}. We then turn to
the FCS in finite temperature equilibrium states, which to the best of
our knowledge has not been considered in the literature.

%%%%%%%%%%%%%%%%%%%%%%%%%%%%%%%%%%%%%%%%%%%%%%%%%%%%%%
\subsection{Full counting statistics in the ground state}
%%%%%%%%%%%%%%%%%%%%%%%%%%%%%%%%%%%%%%%
In the ground state the generating function is of the form \fr{chiTI}, 
\fr{blockTI} with entries
\begin{align}
f_l=&0,\\
g_l=& -i  \int_{-\pi}^{\pi} \frac{d k}{2\pi} e^{-i k l} e^{i\theta_k}\,,
\end{align}
where $\theta_k$ is the Bogoliubov angle \fr{Bogangle}.
By rearranging rows and columns, $\bar{\Gamma}$ can be brought to a block
diagonal form with $\ell\times \ell$ matrices $G$ and $G^T$ \fr{FG} on
the diagonal and zero otherwise. This allows us to express the generating
function as 
\begin{align}
\chi^{(u)}(\lambda,\ell)=& (2\cos\lambda)^\ell 
\sqrt{\det{\frac{1-\tan(\lambda) G}2}\det{\frac{1-\tan(\lambda)
      G^T}2}}= \det{\cos\lambda -\sin(\lambda) G}\ ,
\label{chiGS}
\end{align}
This is precisely the result previously obtained by Cherng and Demler
\cite{cd-07} by a different technique. They considered the generating
function 
\begin{align}
\chi_{CD}(\lambda,\ell)=\langle{\rm GS}| e^{i\lambda \sum_{j=1}^\ell
  \frac{1-\sigma_j^z}2}|{\rm GS}\rangle=e^{i\lambda\ell/2}
\chi^{(u)}(-\lambda/2,\ell)=
\det{\frac{1+e^{i\lambda}}2+ \frac{1-e^{i\lambda}}2 i G}\ .
\end{align}

The Toeplitz determinant \eqref{chiGS} can be analyzed by standard
methods \cite{cd-07}. The \emph{symbol} $\tau(e^{ik})$ of a block
Toeplitz $T_\ell$ with elements $(T_\ell)_{ln}=t_{l-n}$ is defined
through the equation
\begin{align}
t_n = \int_{0}^{2\pi} \frac{dk}{2\pi} \tau(e^{ik}) e^{-ink}.
\label{symbolT}
\end{align}
The symbol of the block Toeplitz matrix \eqref{chiGS} is given by
\begin{align}
\tau(e^{ik})= \cos\lambda+i e^{i\theta_k} \sin\lambda. 
\end{align}
As long as the symbol has zero winding number a straightforward
application of Szeg\H{o}'s Lemma gives, \emph{cf.}
Appendix~\ref{app:Szego}
\begin{align}
\lim_{\ell\to\infty} \frac{\ln\chi^{(u)}(\lambda,\ell)}\ell= 
\int_0^{2\pi} \frac{dk}{2\pi}\ln (\cos\lambda+i e^{i\theta_k}
\sin\lambda)\,. 
\label{eq:szego0w}
\end{align}
For $h<1$ and $\lambda>\lambda_c(h)$ the winding number of the symbol
is $1$ and the above result gets modified accordingly \cite{ia-13}.
For a detailed analysis we refer to Ref.~\cite{ia-13}. The full counting statistics for the transverse magnetization in the entire system was studied in \cite{Eisler-2003} and the result is identical to \eqref{eq:szego0w}. 
Thus considering the subsystem instead of the entire system only makes a difference for $h < 1$ and $\lambda > \lambda_c(h)$, as discussed in \cite{cd-07}.

We note that all cumulants can be obtained from (\ref{eq:szego0w})
since they are defined by the expansion close to $\lambda=0$.
Consequently, the first cumulants are given by
\bea
C_1&=& \int_{-\pi}^\pi \frac{dk}{2\pi}  e^{i\theta_k}, \nn
C_2&=&  \int_{-\pi}^\pi \frac{dk}{2\pi} (1-e^{2i\theta_k}),\nn
C_3&=& \int_{-\pi}^\pi \frac{dk}{2\pi}  2(e^{3i\theta_k}-e^{i\theta_k}) ,\nn
C_4&=& \int_{-\pi}^\pi \frac{dk}{2\pi}  2(-1+4e^{2i\theta_k}-3e^{4i\theta_k}).
\eea
The non-zero values of $C_3$ and $C_4$ show that the probability
distribution is non-gaussian.

%%%%%%%%%%%%%%%%%%%%%%%%%%%%%%%%%%%%%%%%%%%%%%%%%%%%%%%%%%%%%%%%
\subsection{Full counting statistics at finite temperature}
%%%%%%%%%%%%%%%%%%%%%%%%%%%%%%%%%%%%%%%%%%%%%%%%%%%%%%%%%%%%%%%%
We now turn to the FCS in finite temperature equilibrium states, for
which we are not aware of any results in the literature.  
In this case, the correlation matrix has the same structure as for the
ground state, but now
\begin{align}
f_l&=0,\\
g_l&= -i  \int_{-\pi}^{\pi} \frac{d k}{2\pi} e^{-i k l} e^{i\theta_k} \tanh(\beta \epsilon_k/2)\,,
\end{align}
where $\epsilon_k$ is the dispersion relation \fr{dispersion}. 
Since $f_l=0$, the same simplifications as in the ground state case
apply and the generating function can be expressed as
\begin{align}
\chi^{(u)}(\lambda,\ell)= \det{\cos\lambda -\sin(\lambda) G}.
\label{chiT}
\end{align}
\begin{figure}[ht!]
(a)\includegraphics[width=0.45\textwidth]{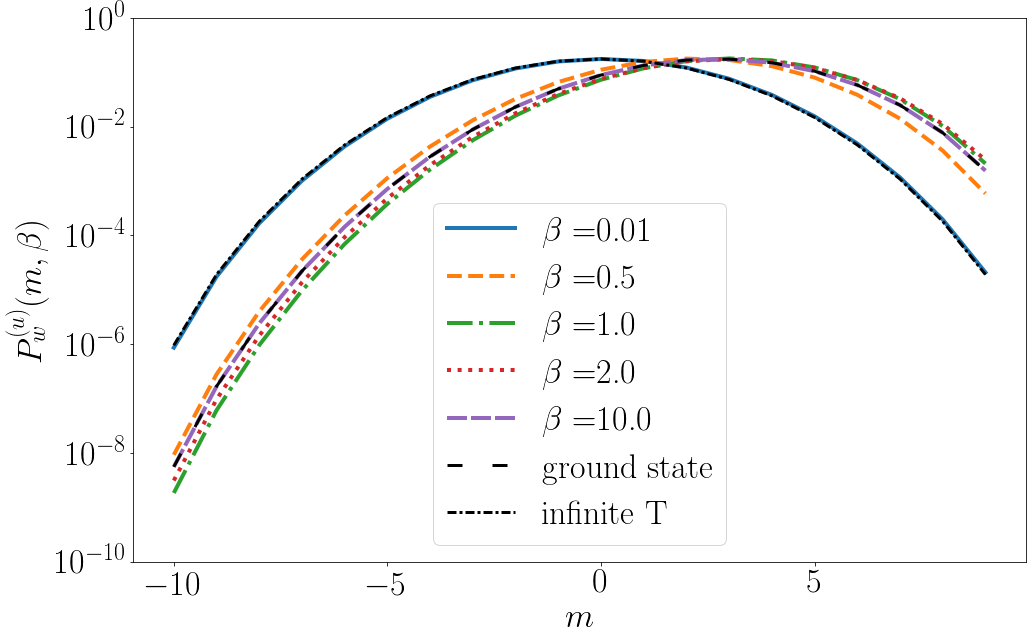}
\qquad
(b)\includegraphics[width=0.45\textwidth]{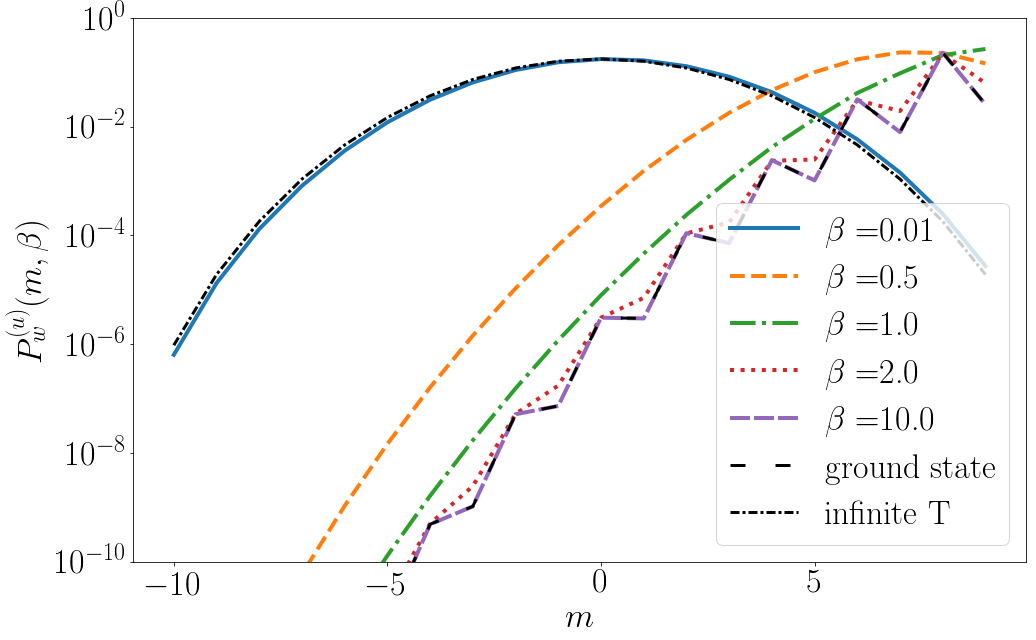} 
\caption{Probability distribution as a function of $m$ for $\ell=20$
and several temperatures at (a) $h=0.5$; (b) $h=2$.}
\label{Fig:finiteT}
\end{figure}
In Fig.~\ref{Fig:finiteT} we show $P_w^{(u)}(m)$ for subsystem size
$\ell=20$ and several different temperatures. We employ a log-linear
plot in order to make the deviations of the probability distributions
from a Gaussian form (which would correspond to a parabolic form) more
apparent. We can see from Fig.~\ref{Fig:finiteT} (a) that the
temperature dependence for $h<1$, corresponding to the
ferromagnetically ordered phase at zero temperature, is not very
pronounced. In contrast we see a much stronger temperature dependence
in the paramagnetic phase, \emph{cf.} Fig.~\ref{Fig:finiteT} (b). At
low temperatures the probability distribution is as expected
asymmetric as a result of the applied field and is seen to display an
even/odd structure. The latter disappears quickly as temperature is
increased, whereas the asymmetry remains until the temperature exceeds
the scale set by the magnetic field.
\begin{figure}[ht!]
(a)\includegraphics[width=0.45\textwidth]{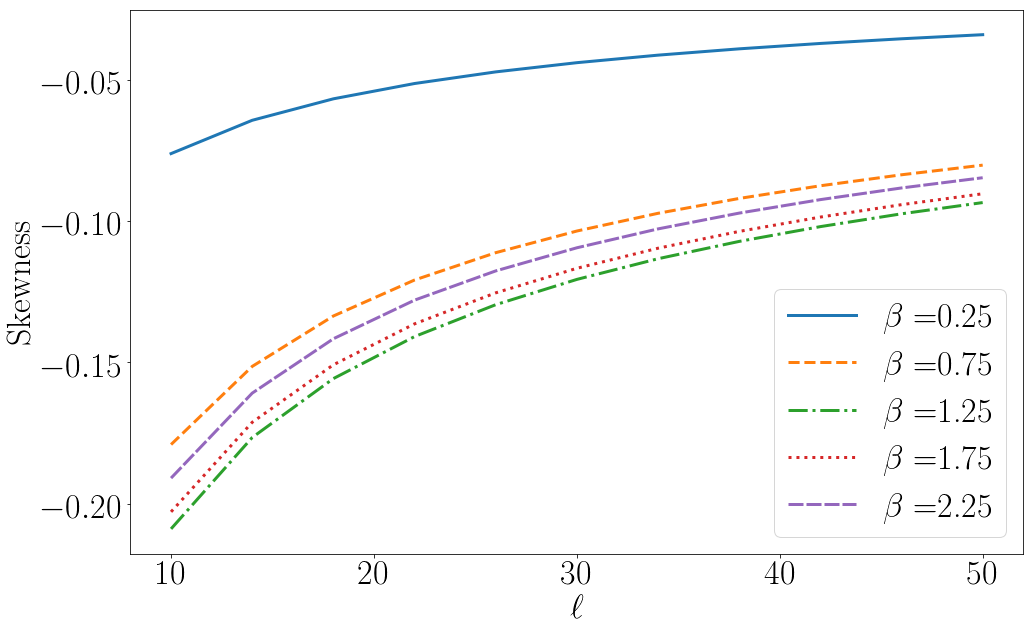}
\qquad
(b)\includegraphics[width=0.45\textwidth]{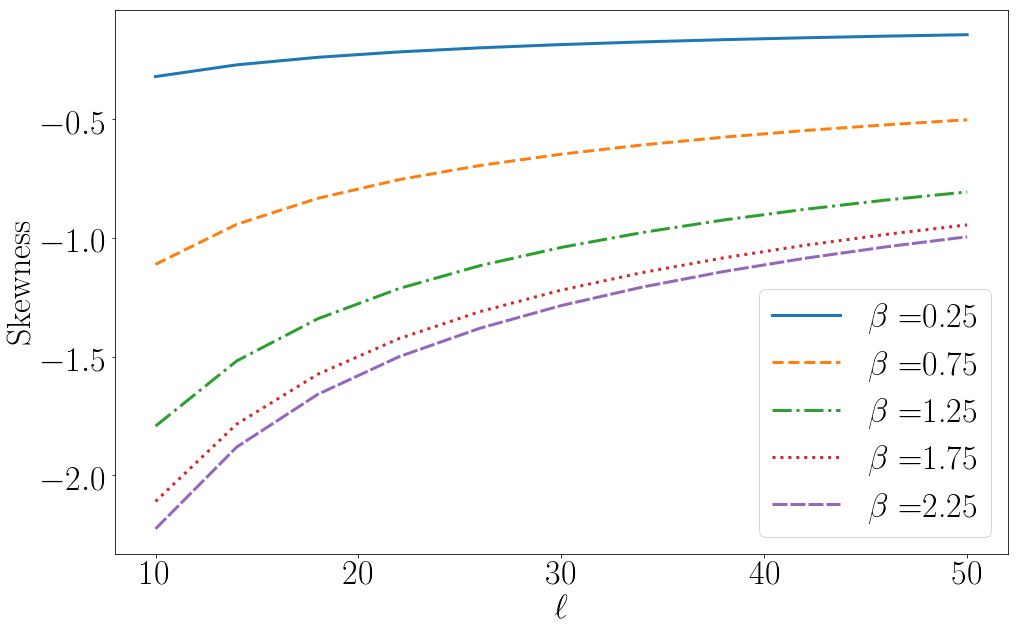}
\caption{Skewness as a function of $\ell$ for several values of
  $\beta$ at (a) $h=0.5$ and (b) $h=2$.}
\label{Fig:finiteT2}
\end{figure}

\begin{figure}[ht!]
(a)\includegraphics[width=0.45\textwidth]{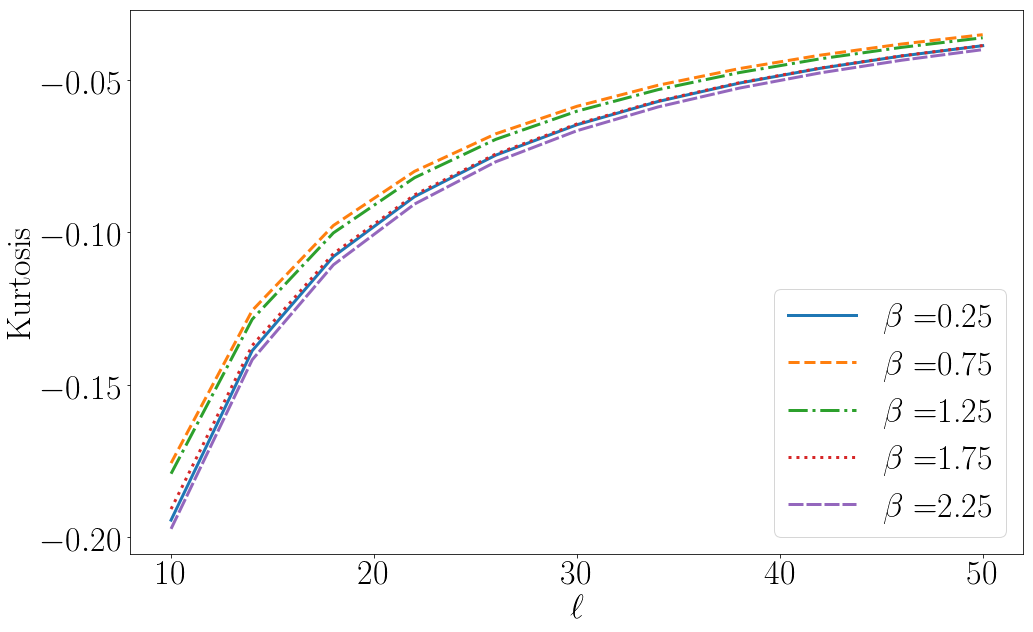}
\qquad
(b)\includegraphics[width=0.45\textwidth]{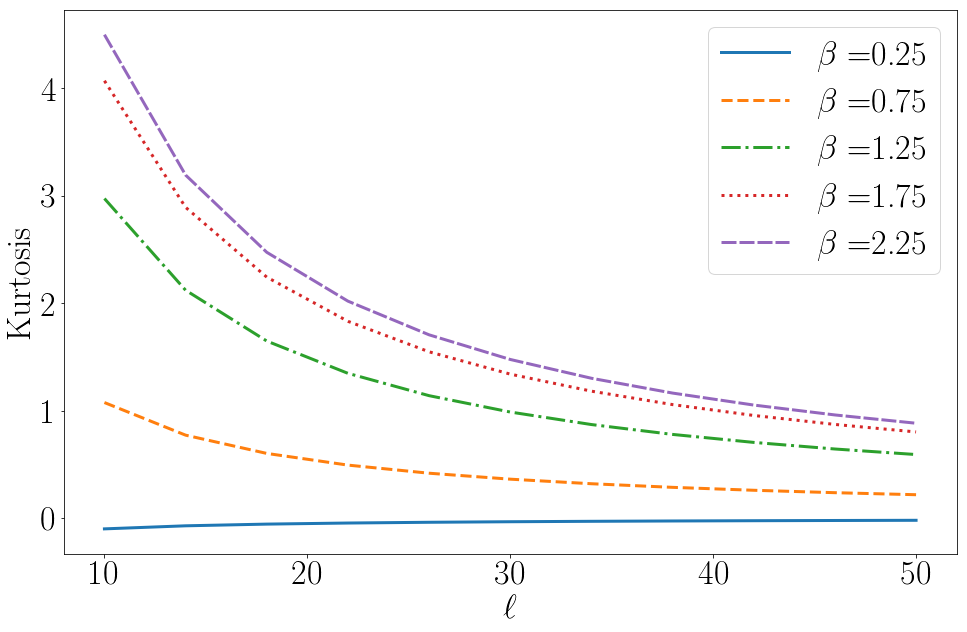}
\caption{Excess kurtosis as a function of subsystem size $\ell$ for several
temperatures and (a) $h=0.5$ and (b) $h=2$.}
\label{Fig:finiteT3}
\end{figure}
In Figs~\ref{Fig:finiteT2} and \ref{Fig:finiteT3} we show the skewness
and excess kurtosis of the probability distribution as a function of
subsystem size $\ell$ for a range of temperatures. These are defined
as the thermal expectation values
\be
\Big\langle \left[\frac{X}{\sqrt{\langle X^2\rangle_\beta}}\right]^3
\Big\rangle_\beta\ ,\quad
\Big\langle \left[\frac{X}{\sqrt{\langle X^2\rangle_\beta}}\right]^4
\Big\rangle_\beta-3\ ,\quad
X=S^z_u(\ell)-\langle S^z_u(\ell)\rangle_\beta\ .
\ee
Both skewness and excess kurtosis are non-vanishing for finite $\beta$
and $\ell$, which establishes that the distribution is not Gaussian.
A very peculiar feature is that at fixed $\ell$ skewness and excess
kurtosis are non-monotonic functions of the temperature. Furthermore,
we observe that at a fixed temperature they both tend to zero as the
subsystem size $\ell$ is increased. This signals that the
corresponding probability distribution approaches a Gaussian. This 
is expected as for large subsystem sizes the laws of thermodynamics
apply and the probability distribution is then approximately Gaussian
with a standard deviation that scales as $\sqrt{\ell}$.
%%%%%%%%%%%%%%%%%%%%%%%%%%%%%%%%%%%%%%%%%%%%%%%%%%%%%%%%%%%
\section{Full counting statistics after a quantum quench}
\label{sec:QQ}
%%%%%%%%%%%%%%%%%%%%%%%%%%%%%%%%%%%%%%%%%%%%%%%%%%%%%%%%%%%
We now turn to the time evolution of the characteristic function
$\chi^{(u,s)}(\lambda,t)$ after quantum quenches. We consider two different
classes of initial states:
\begin{itemize}
\item{} We initialize the system in the ground state of $H(h_0)$ and
time evolve with $H(h)$. Such transverse field quenches have been
studied in detail in the literature \cite{IR:quench00,sps-04,R1,R2,CEF1,CEF2,CEF3,cc-05,FC08,FE_13a,EEF:12,FCG:f-d11,RI:sc11,SE:Ising,FCG:dyn12,HPL:dyn13,kbc-14,bkc-14,ak-16}.
\item{} We initialize the system in the N\'eel state
$\ket{\uparrow\downarrow\uparrow\downarrow\dots\uparrow\downarrow}$, thus
breaking translational symmetry by one site. This symmetry is restored
at late times after the quench and it is an interesting question how
this is reflected in the probability distributions of observables.
\end{itemize}
%%%%%%%%%%%%%%%%%%%%%%%%%%%%%%%%%%%%%%%%%%%%%%%%%%%%%%%%%%%%%
\subsection{Transverse field quench $h_0\longrightarrow h$}
%%%%%%%%%%%%%%%%%%%%%%%%%%%%%%%%%%%%%%%%%%%%%%%%%%%%%%%%%%%%%
In this quench protocol both the Hamiltonian and the initial state 
are translationally invariant. The characteristic function has the
determinant representation \fr{chiTI}, \fr{blockTI} with \cite{cc-05}
\begin{align}
g_l &= -i \int_{-\pi}^\pi \frac{\id k}{2\pi} e^{-ikl}e^{i\theta_k}
\left(\cos\Delta_k - i \sin\Delta_k \cos(2\eps_k t)\right) \\
f_l &= \int_{-\pi}^\pi \frac{\id k}{2\pi} e^{-ikl}\sin\Delta_k
\sin(2\eps_k t)\ ,
\end{align}
where 
\be
e^{i\theta_k} = \frac{h-e^{ik}}{\sqrt{1+h^2-2h\cos k}}\ ,\quad
\cos\Delta_k=4\frac{hh_0 - (h+h_0)\cos k
  +1}{\eps_h(k)\eps_{h_0}(k)}\ .
\label{thetaDelta}
\ee
Using Szeg\H{o}'s Lemma it is straightforward to obtain the
large-$\ell$ asymptotics in the initial ($t=0$) and stationary
($t=\infty$) states. The $t=0$ result corresponds to a ground state at
field $h_0$ and has been discussed earlier. 
%%%%%%%%%%%%%%%%%%%%%%%%%%%%%%%%%%%%%%%%%%%%%%%%%%%%%%
\subsubsection{Behaviour in the stationary state}\label{sec:stationarystate}
%%%%%%%%%%%%%%%%%%%%%%%%%%%%%%%%%%%%%%%%%%%%%%%%%%%%%%
The late time asymptotics of the generating function can be determined
from Szeg\H{o}'s Lemma. For quenches into the paramagnetic phase $h>1$
it takes the form
\begin{align}
\lim_{t\to\infty}
\frac{\ln\chi^{(u)}(\lambda,\ell,t)}{\ell}=
\int_0^{2\pi} \frac{dk}{2\pi} \ln\left( \cos\lambda +
i\sin\lambda\cos\Delta_ke^{i\theta_k}\right)+
\mathcal{O}(1/\ell)\ ,\quad
\ell\gg 1.
\label{statstate}
\end{align}
The ${\cal O}(\ell^{-1})$ corrections also follow from Szeg\H{o}'s
Lemma.
\begin{figure}[ht]
(a)\includegraphics[width=0.45\textwidth]{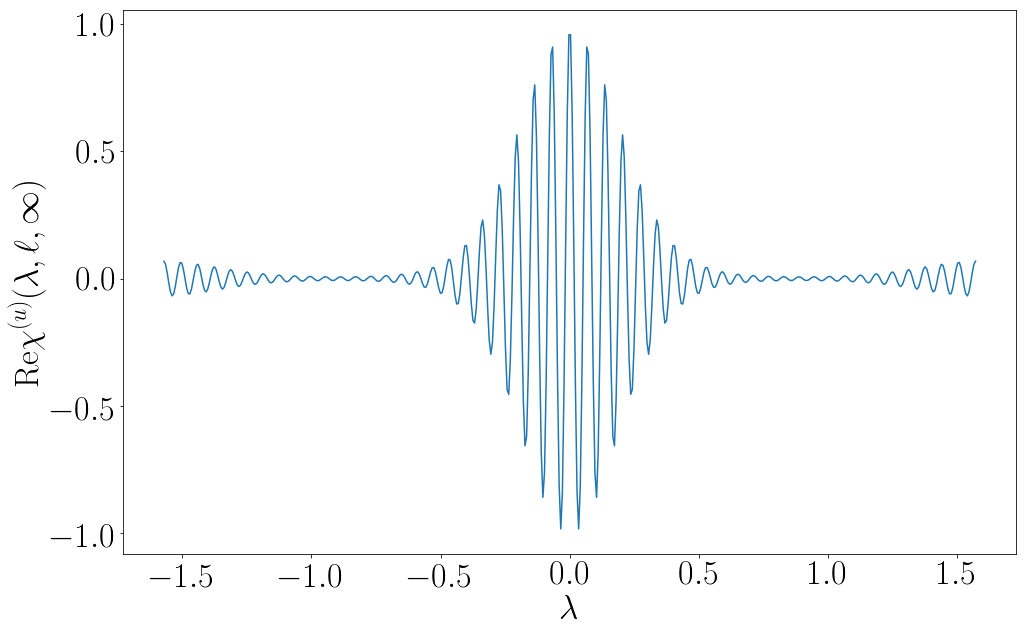}
\qquad
(b)\includegraphics[width=0.45\textwidth]{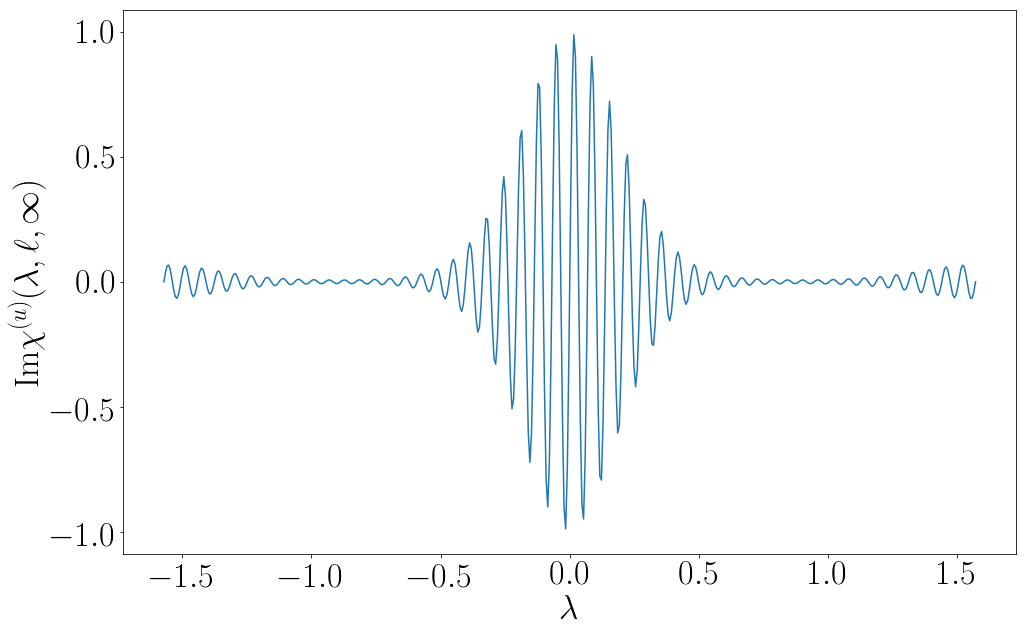}
\caption{(a) ${\rm Re}\chi^{(u)}(\lambda,\ell,\infty)$ and
${\rm Im}\chi^{(u)}(\lambda,\ell,\infty)$ for a
  quench from $h=5$ to $h=2$ and subsystem size $\ell=100$.}
\label{Fig:chisteady}
\end{figure}
The real and imaginary parts of $\chi^{(u)}(\lambda,\ell,t)$
(with ${\cal O}(\ell^{-1})$ corrections included) are shown for a
transverse field quench from $h_0=5$ to $h=2$ and subsystem size
$\ell=100$ in Fig.~\ref{Fig:chisteady}.  

\begin{figure}[ht!]
\includegraphics[width=0.45\textwidth]{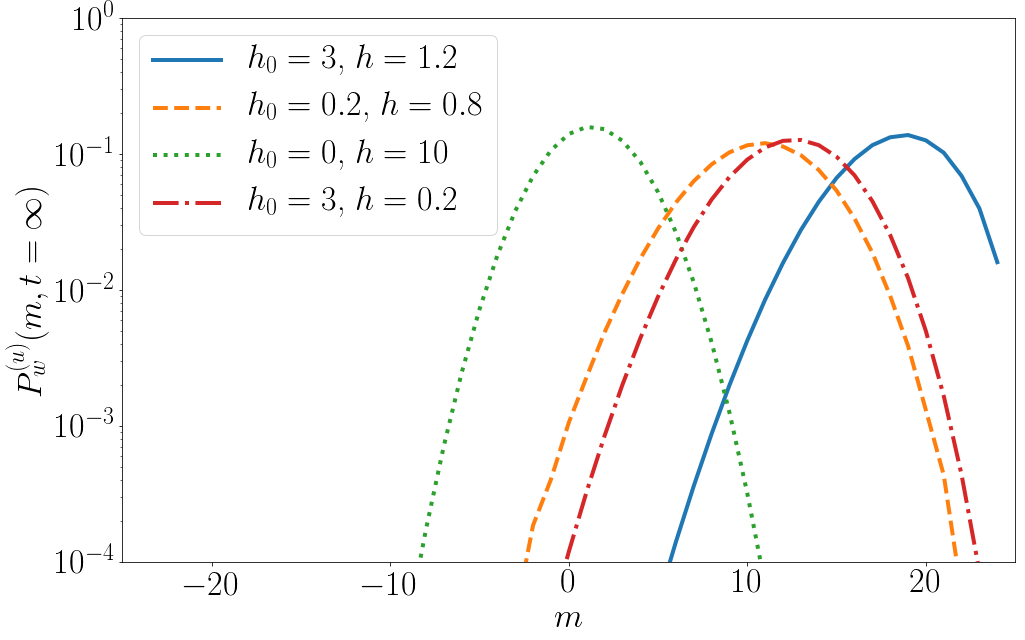}
\caption{Stationary state probability distribution
$P^{(u)}_w(m,\infty)$ for a subsystem of size $\ell=70$ for several
transverse field quenches.} 
\label{Fig:probsteady}
\end{figure}

\begin{figure}[ht!]
(a)\includegraphics[width=0.45\textwidth]{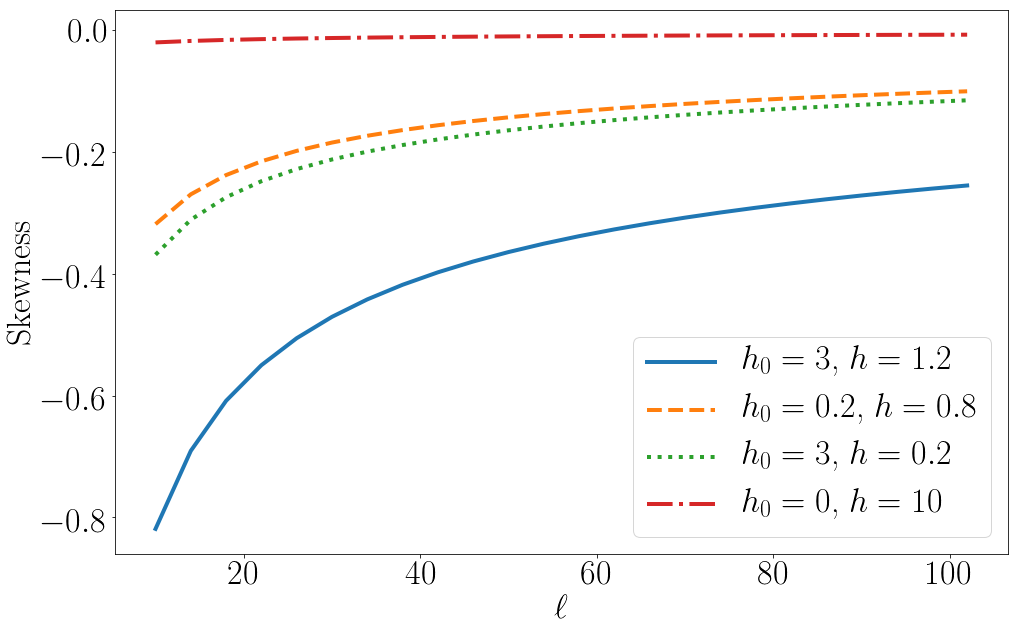}
\qquad
(b)\includegraphics[width=0.45\textwidth]{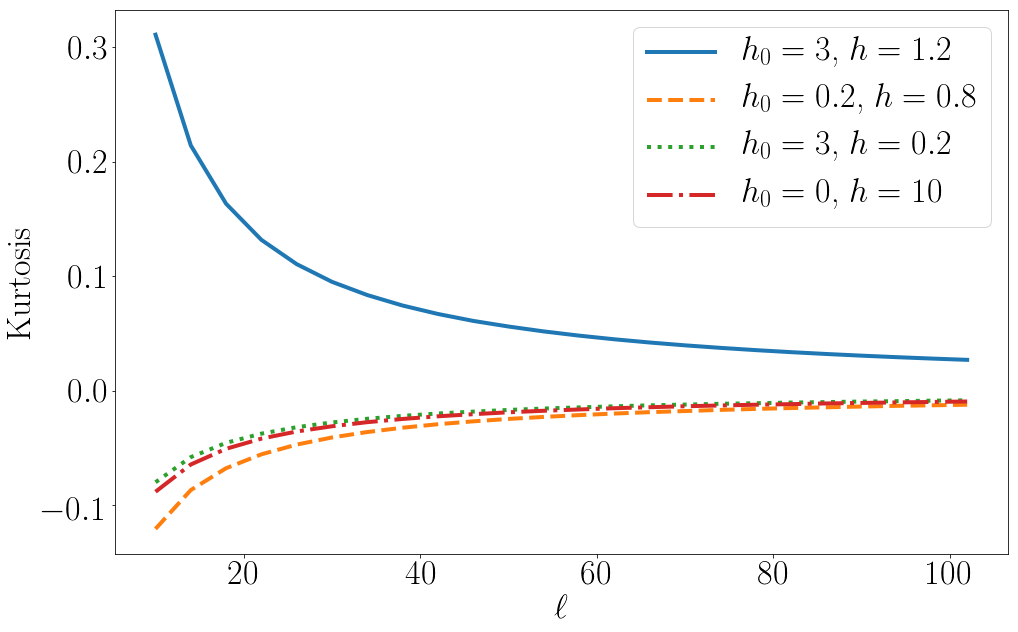}
\caption{(a) Skewness and (b) Excess kurtosis of the steady state
probability distribution as functions of subsystem size $\ell$
for a number of transverse field quenches.}
\label{Fig:probsteady_2}
\end{figure}

For quenches into the ferromagnetic phase and
$\lambda<\lambda_c(h_0,h)$, Eq.  \fr{statstate} continues to hold. However, for
$\lambda>\lambda_c(h_0,h)$ the symbol exhibits non-zero winding number and
the analysis needs to be modified, \emph{cf.} Appendix \ref{app:Szego}. The probability distribution in the stationary state is obtained by
Fourier transforming $\chi^{(u)}(\lambda,\ell,t)$. Examples for several
transverse field quenches are shown in Fig.~\ref{Fig:probsteady}. We
again employ a logarithmic scale to make the deviations from a
Gaussian form more apparent. 
In Figs~\ref{Fig:probsteady_2} we plot
the skewness and the excess kurtosis of the steady state probability
distributions for a number of transverse field quenches.
We observe that in all cases both skewness and excess kurtosis tend to
zero for large subsystem sizes. This signals that the probability
distributions approach Gaussians in the large-$\ell$ limit. While the
steady states are non-thermal now, they still exhibit finite
correlation lengths. Employing the same arguments as for finite
temperature ensembles then implies that the cumulants of
$S^z_u(\ell)$ are proportional to $\ell$ in the large-$\ell$
limit. This in turn suggests that skewness and excess kurtosis should
scale as $\ell^{-1/2}$ and $\ell^{-1}$ respectively, while the
standard deviation scales as $\ell^{1/2}$. These expectations are in
perfect agreement with our findings.
%%%%%%%%%%%%%%%%%%%%%%%%%%%%%%%%%%%%%%
\subsubsection{Scaling collapse}
\label{sec:scaling}
\begin{figure}[ht!]
(a)\includegraphics[width=0.45\textwidth]{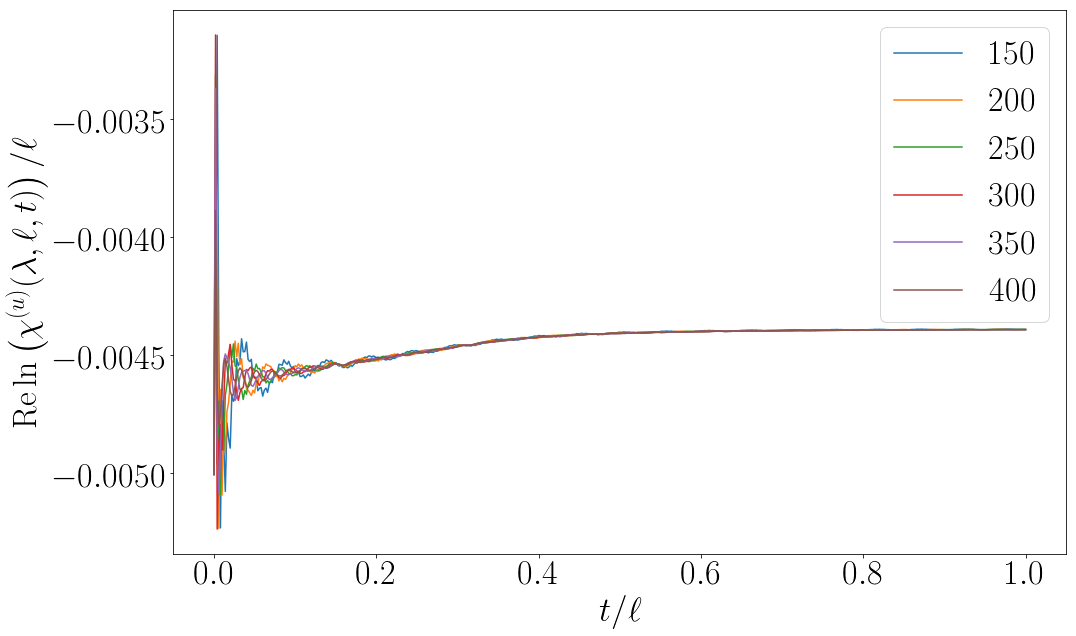}
\qquad
(b)\includegraphics[width=0.45\textwidth]{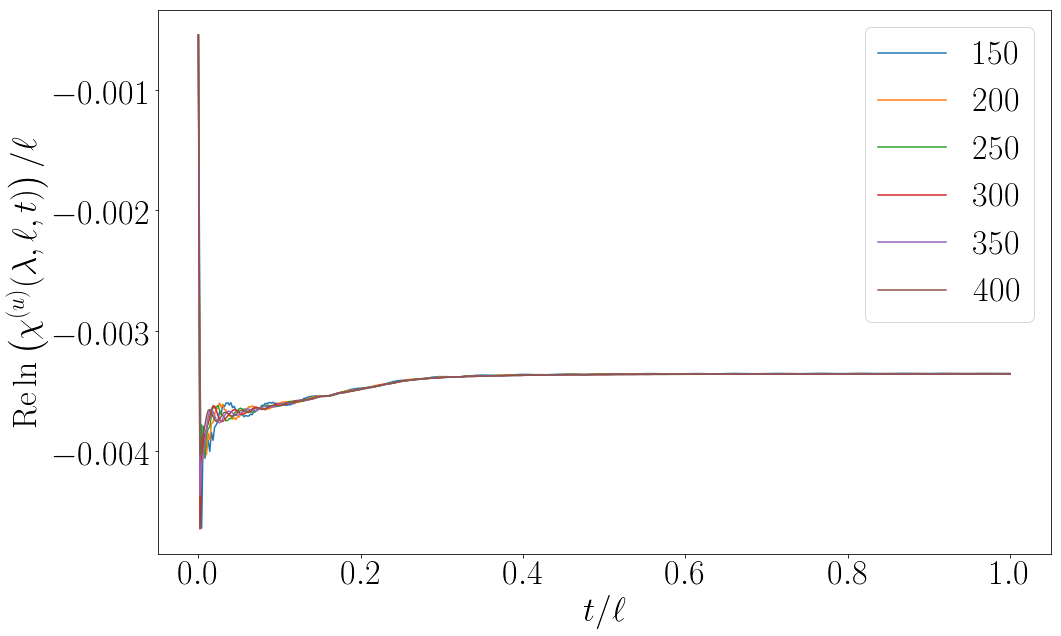}
\caption{${\rm Re}\ln\chi^{(u)}(0.1,\ell,t)/\ell$ for several values of $\ell$ for a
quench from (a) $h=0.2$ to $h=0.8$ and (b) $h=3$ to $h=1.2$. The data
for different subsystem sizes are seen to collapse at sufficiently late times.}
\label{Fig:logchi_collapse1}
\end{figure}

\begin{figure}[ht!]
(a)\includegraphics[width=0.45\textwidth]{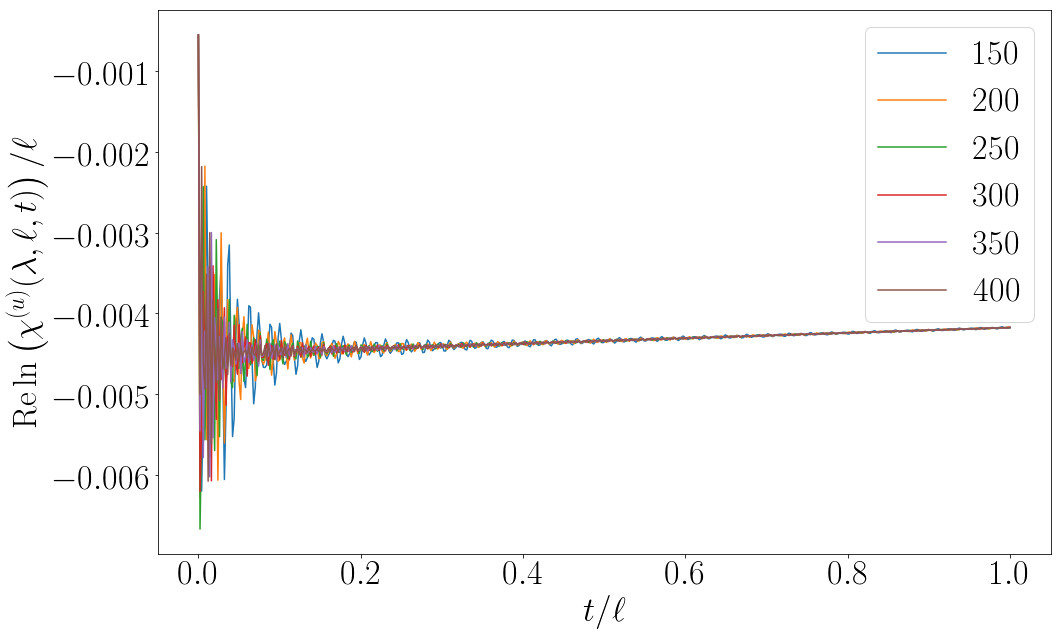}
\qquad
(b)\includegraphics[width=0.45\textwidth]{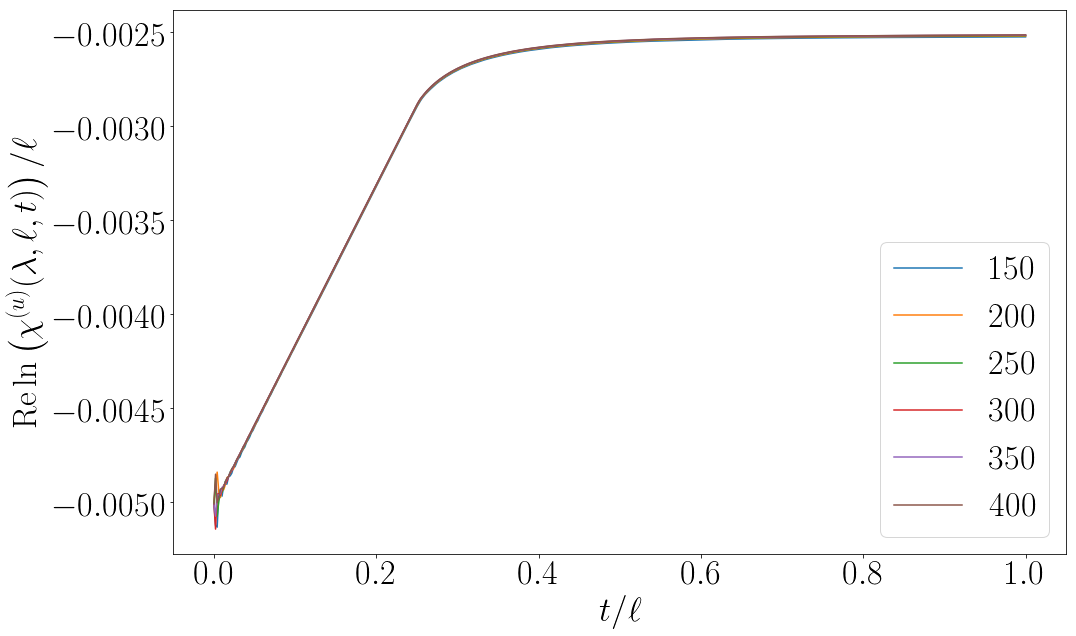}
\caption{
${\rm Re}\ln\chi^{(u)}(0.1,\ell,t)/\ell$ for several values of $\ell$ for a
quench from (a) $h=3$ to $h=0.2$ and (b) $h=0$ to $h=20$. The data
for different subsystem sizes are seen to collapse at sufficiently late times.}
\label{Fig:logchi_collapse2}
\end{figure}

%%%%%%%%%%%%%%%%%%%%%%%%%%%%%%%%%%%%%%
At finite times the FCS and the probability distribution can be
computed efficiently from the determinant representation \fr{chiTI}.
Importantly we observe that for sufficiently large values of
$\ell$ and $t$ there is scaling collapse
\be
\chi^{(u)}(\lambda,\ell,t)\approx {\rm exp}\big(\ell f(t/\ell)\big)\ ,\quad t,\ell\gg 1.
\label{scaling}
\ee
The property \fr{scaling} is an important ingredient in the analytic
calculation of the FCS described in section
\ref{sec:analytic}. Several examples of the scaling behaviour of
the real part of the generating function are shown in
Figs~\ref{Fig:logchi_collapse1}, \ref{Fig:logchi_collapse2}. The
imaginary parts exhibit a similar scaling collapse.  

\begin{figure}[ht!]
\includegraphics[width=0.45\textwidth]{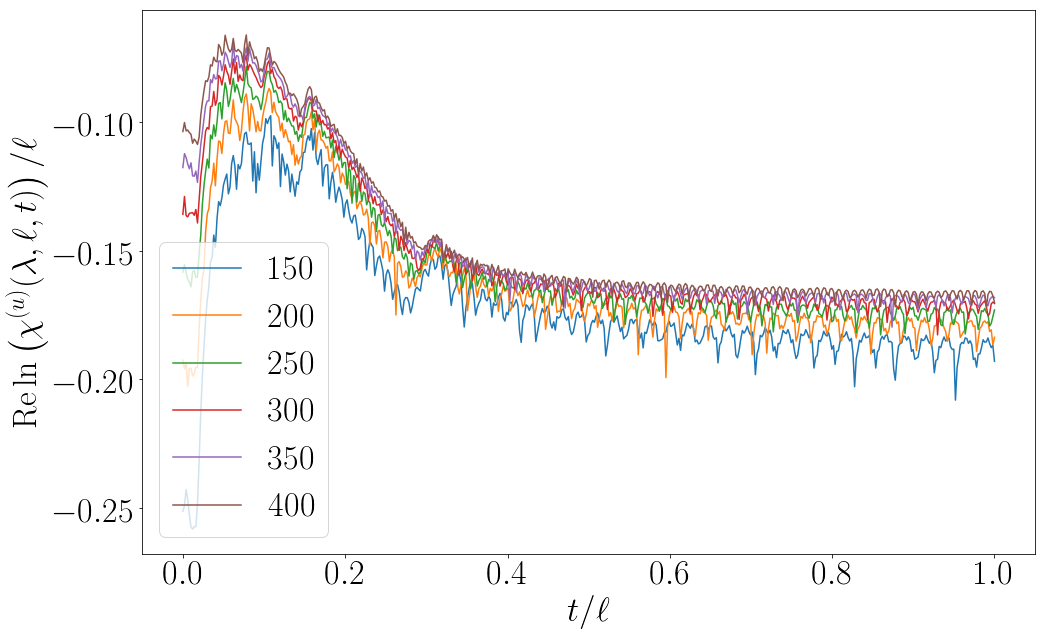}
\qquad
\caption{${\rm Re}\ln\chi^{(u)}(1.4,\ell,t)/\ell$ for several values
of $\ell$ for a quench from $h=0.2$ to $h=0.8$. The data
for different subsystem sizes are seen to collapse at sufficiently
late times only for very large subsystem sizes $\ell$.} 
\label{Fig:logchi_collapse_largela}
\end{figure}

For quenches \emph{towards} the ferromagnetic regime the scaling
collapse for general values of $\lambda$ can be significantly worse,
and then really only emerges at rather large subsystem sizes $\ell$,
\emph{cf.} Fig.~\ref{Fig:logchi_collapse_largela}. 
Like in the case of the stationary state discussed above, \emph{c.f.}
section \ref{sec:stationarystate}, there exists a critical value
$\hat\lambda_c(h_0,h)$ of the counting parameter such that for 
$\lambda<\hat\lambda_c(h_0,h)$ the scaling collapse is excellent, while
for $\lambda>\hat\lambda_c(h_0,h)$ no collapse is observed at the
times and subsystem sizes of interest here. For the cases we have
considered $\hat\lambda_c(h_0,h)$ coincides with $\lambda_c(h_0,h)$,
which is the value of the counting parameter above which the symbol
has non-zero winding number.
We note however, that in cases like the one shown in
Fig.~\ref{Fig:logchi_collapse_largela} the generating function itself
is extremely small and will not give a significant contribution to the
corresponding probability distribution. 

%%%%%%%%%%%%%%%%%%%%%%%%%%%%%%%%%%%%%%%%%%%%%%%%%%%%%%%%%%%%%%%%%%%%
\subsubsection{Time dependence of the probability distribution}
%%%%%%%%%%%%%%%%%%%%%%%%%%%%%%%%%%%%%%%%%%%%%%%%%%%%%%%%%%%%%%%%%%%%
There are basically four different kinds of transverse field quenches
and we now consider them in turn.
\begin{enumerate}
\item{Quenches within the ferromagnetic phase.}
\begin{figure}[ht!]
(a)\includegraphics[width=0.45\textwidth]{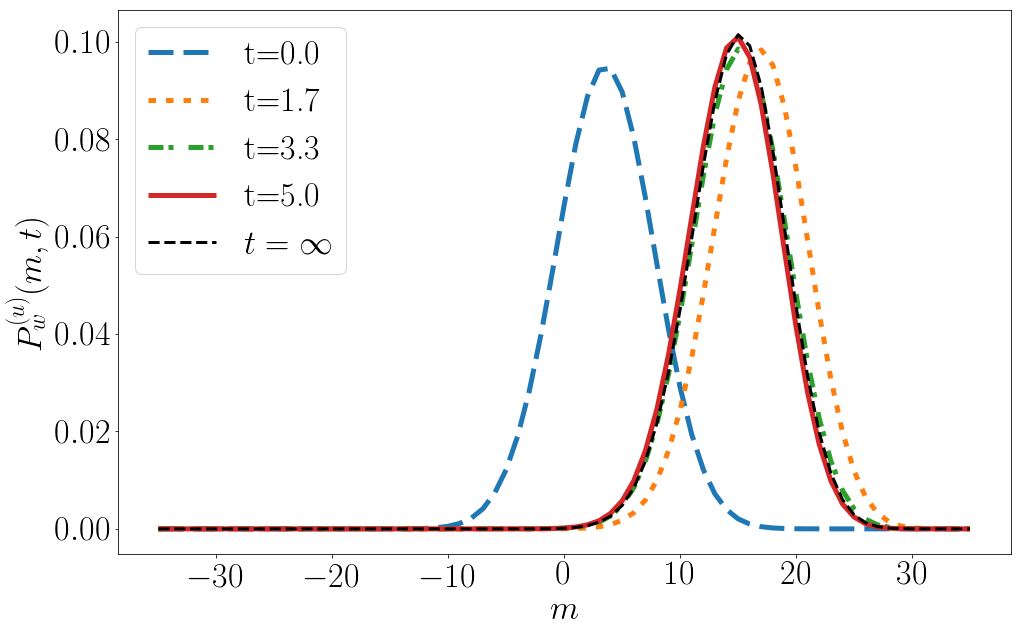}
\qquad
(b)\includegraphics[width=0.45 \textwidth]{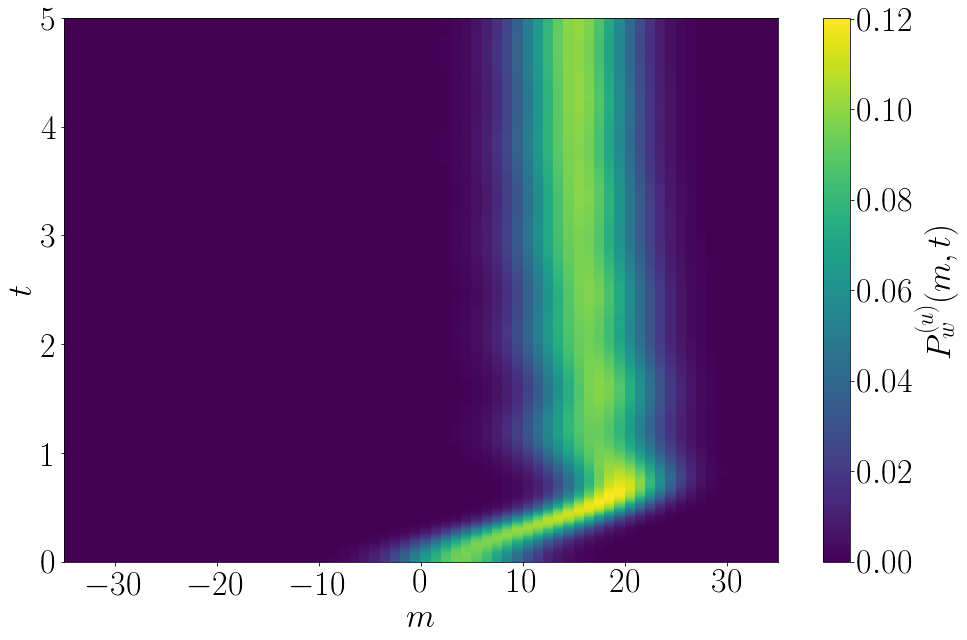}
\caption{(a) Probability distribution $P^{(u)}_w(m,t)$ at times
$t=0,1.7,3.3,5.0$ after a quench from  $h=0.2$ to $h=0.8$ for
subsystem size $\ell=70$. (b) Probability distribution $P^{(u)}_w(m,t)$
for the same parameters.}
\label{fig:FtoF}
\end{figure}
For such quenches the probability distribution remains very narrow and
approximately Gaussian throughout, \emph{cf.} Fig.~\ref{fig:FtoF}. For the
parameters considered the average relaxes quickly towards its
stationary value.

%%%%%%%%%%%%%%%%%%%%%%%%%%%%%%%%%%%%%%%%%%%%%%%%%%%%%%%%%%%%%%%%%
\item{Quenches within the paramagnetic phase.}
%%%%%%%%%%%%%%%%%%%%%%%%%%%%%%%%%%%%%%%%%%%%%%%%%%%%%%%%%%%%%%%%%

\begin{figure}[ht!]
(a)\includegraphics[width=0.45\textwidth]{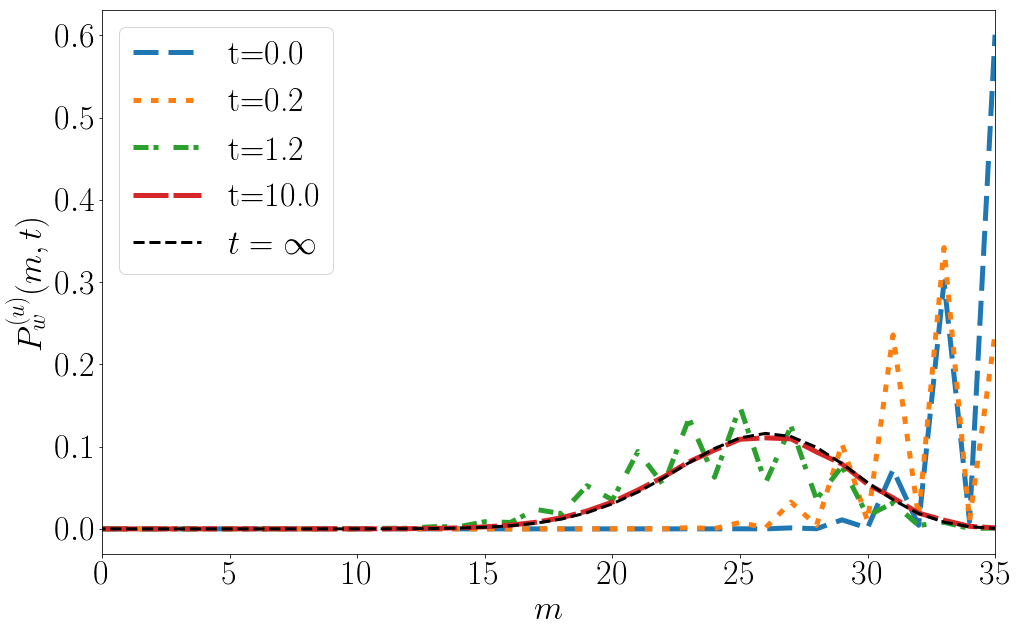}
\qquad
(b)\includegraphics[width=0.45 \textwidth]{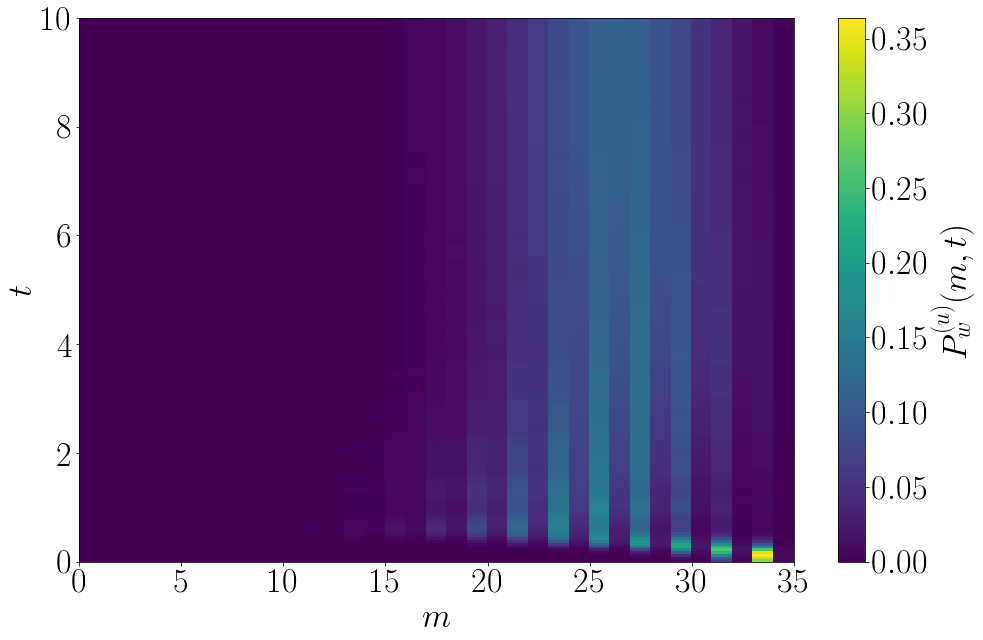}
\caption{(a) Probability distribution $P^{(u)}_w(m,t)$ at times
$t=0,0.2,1.2,10.0$ after a quench from  $h=3$ to $h=1.2$ for
subsystem size $\ell=70$. (b) Probability distribution $P^{(u)}_w(m,t)$
for the same parameters.}
\label{fig:PtoP}
\end{figure}
Here the initial probability distribution exhibits an even/odd structure.
This can be understood by doing perturbation theory around the large
$h_0$ limit, \emph{cf.} Appendix \ref{appendix:perturb}. After the
quench the mean of the probability distribution broadens and shifts
towards smaller values of $m$. The alternating structure is initially
preserved but then gets smoothed out. At late times $P^{(u)}_w(m,t)$
is well described by a Gaussian.

%%%%%%%%%%%%%%%%%%%%%%%%%%%%%%%%%%%%%%%%%%%%%%%%%%%%%%%%%%%%%%%%%
\item{Quenches from the paramagnetic to the ferromagnetic phase.}
%%%%%%%%%%%%%%%%%%%%%%%%%%%%%%%%%%%%%%%%%%%%%%%%%%%%%%%%%%%%%%%%%

\begin{figure}[ht!]
(a)\includegraphics[width=0.45\textwidth]{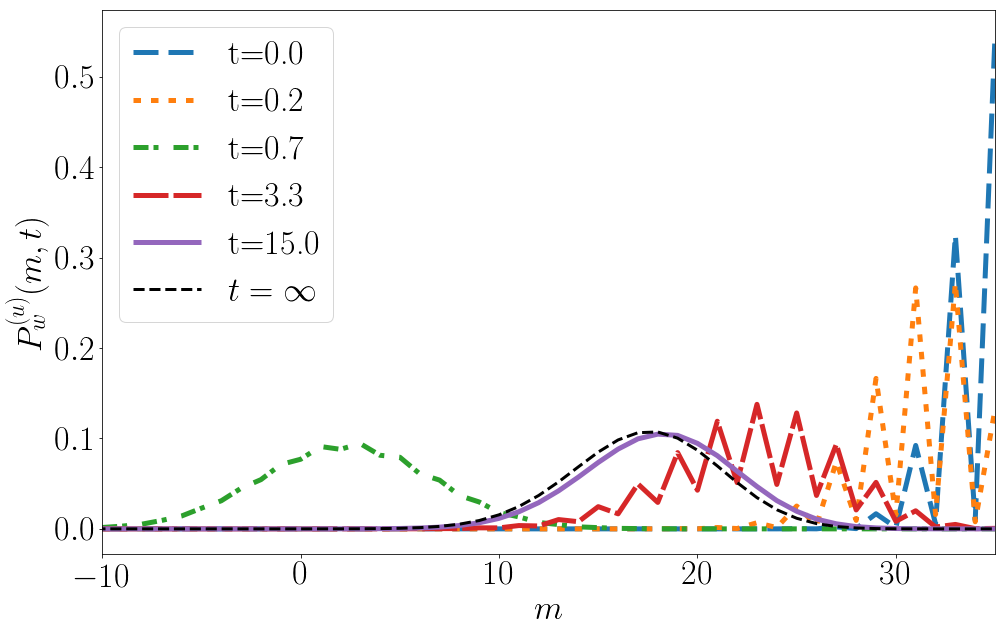}
\qquad
(b)\includegraphics[width=0.45 \textwidth]{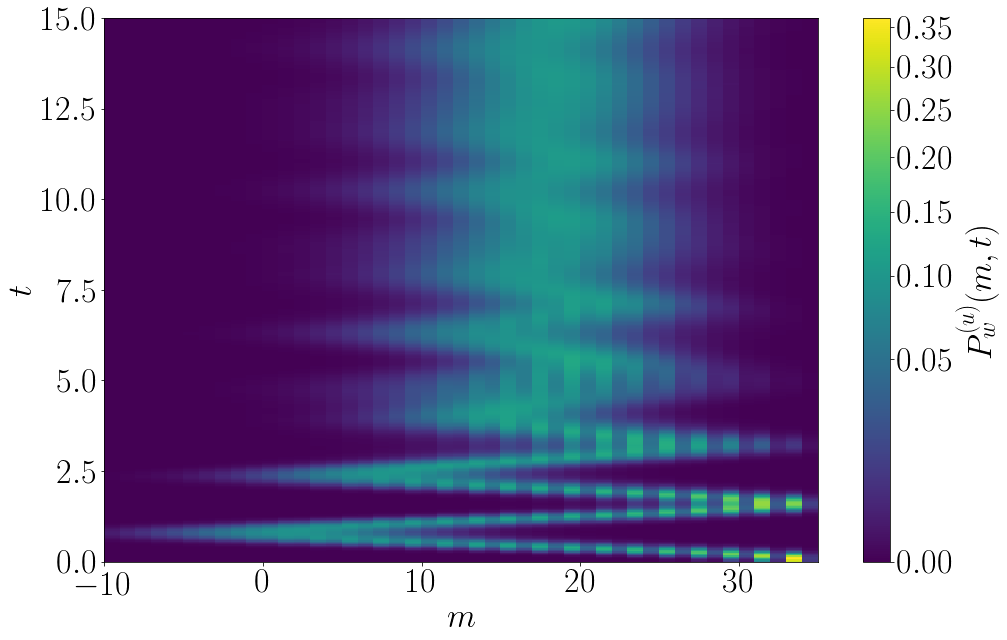}
\caption{(a) Probability distribution $P^{(u)}_w(m,t)$ at times
$t=0,0.2,0.7,3.3,15.0$ after a quench from  $h=3$ to $h=0.2$ for
subsystem size $\ell=70$. (b) Probability distribution $P^{(u)}_w(m,t)$
for the same parameters.}
\label{fig:PtoF}
\end{figure}
Here the probability distribution is initially peaked at a large value
of $m$ and displays an even/odd structure. At later times it broadens
and becomes smooth, while relaxing towards its stationary profile in an
strongly oscillatory manner.
%%%%%%%%%%%%%%%%%%%%%%%%%%%%%%%%%%%%%%%%%%%%%%%%%%%%%%%%%%%%%%%%%
\item{Quenches from the ferromagnetic to the paramagnetic phase.}
%%%%%%%%%%%%%%%%%%%%%%%%%%%%%%%%%%%%%%%%%%%%%%%%%%%%%%%%%%%%%%%%%

\begin{figure}[ht!]
(a)\includegraphics[width=0.45\textwidth]{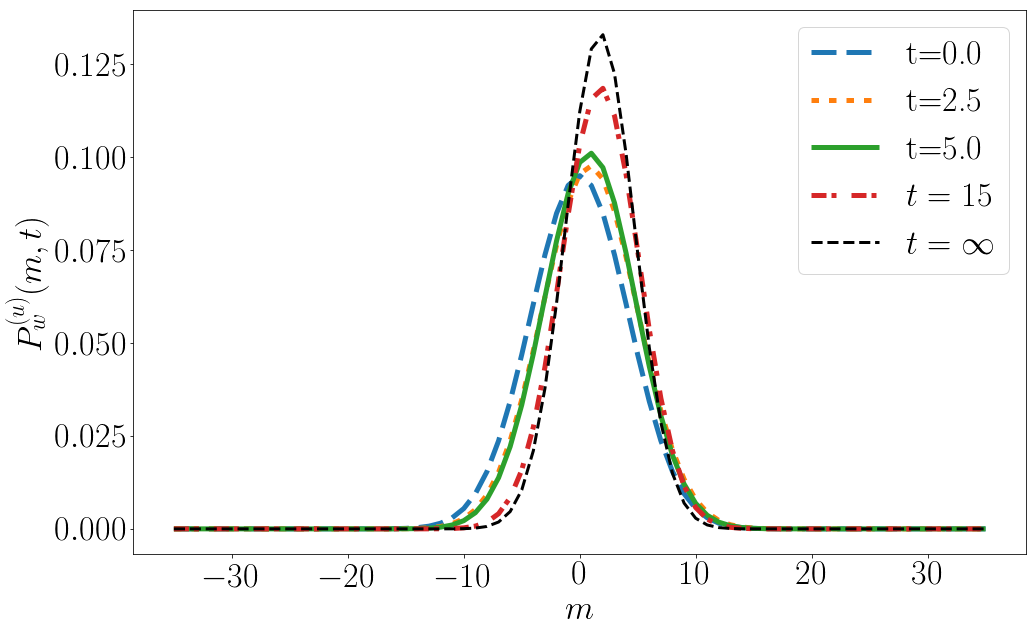}
\qquad
(b)\includegraphics[width=0.45\textwidth]{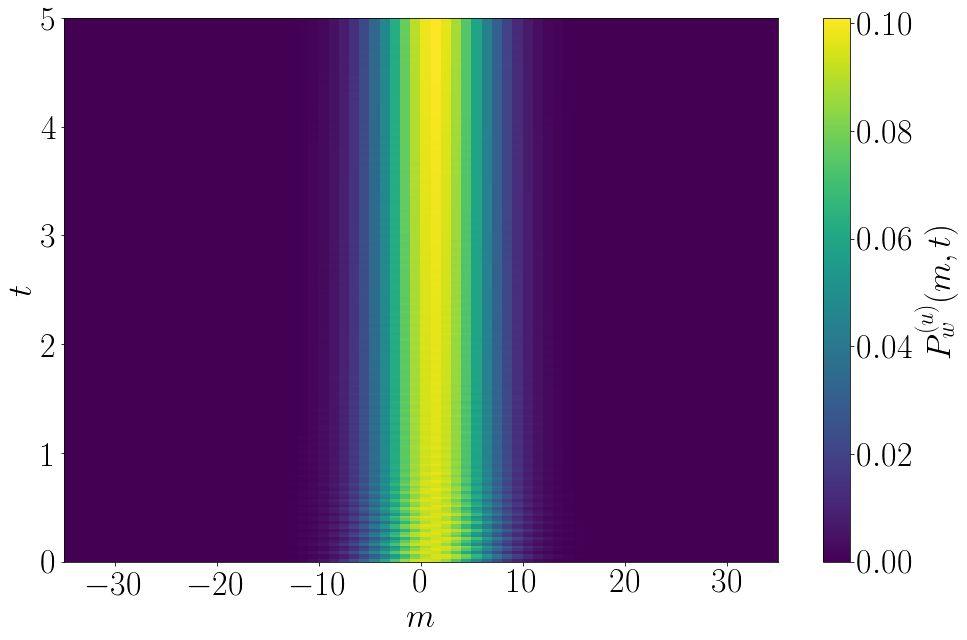}
\caption{(a) Probability distribution $P^{(u)}_w(m,t)$ at times
$t=0,2.5,5.0$ after a quench from  $h=0$ to $h=10$ for
subsystem size $\ell=70$. (b) Probability distribution $P^{(u)}_w(m,t)$
for the same parameters.}
\label{fig:FtoP}
\end{figure}
In this case the probability distribution shows very little variation
in time and remains narrow and approximately Gaussian throughout the
evolution. It was pointed out in Ref.~\cite{HPL:dyn13} that the return
amplitude exhibits a non-analyticity at some finite time $t^*$ after the
quantum quench. This phenomenon was termed a ``dynamical phase
transition''. Local operators are known to be insensitive to this
phenomenon \cite{CEF1,CEF2,CEF3}. We have investigated the behaviour of
$P^{(u)}_w(m,t)$ in the vicinity of $t^*$ but have not observed any unusual
effects. We conclude that the probability distribution for the smooth subsystem magnetization in the transverse field direction is also
insensitive to the ``dynamical phase transition''.
\end{enumerate}

%%%%%%%%%%%%%%%%%%%%%%%%%%%%%%%%%%%%%%%%%%%%%%%%%%%%%%%%%
\subsection{Quench from the N\'eel state}
%%%%%%%%%%%%%%%%%%%%%%%%%%%%%%%%%%%%%%%%%%%%%%%%%%%%%%%%%
We now turn to the time evolution of $P^{(u,s)}_w(m,t)$ when the system is
initialized in the N\'eel state 
$\ket{\psi_0}=\ket{\uparrow\downarrow\uparrow\downarrow\dots
\uparrow\downarrow}$. This explicitly breaks translational invariance
by one site, but retains invariance under translation by two
sites. As a result the subsystem correlation matrix is now a $4\times 4$
block-Toeplitz matrix 
\begin{align}
\Gamma^A_{\text{N\'eel}}=\begin{pmatrix} 
\Pi^{\text{N\'eel}}_0 & \Pi^{\text{N\'eel}}_{-1} & \dots & \Pi^{\text{N\'eel}}_{1-\ell/2} \\
\Pi^{\text{N\'eel}}_1 & \Pi^{\text{N\'eel}}_0 & & \vdots \\
    \vdots & & \ddots & \vdots \\
    \Pi^{\text{N\'eel}}_{\ell/2-1} &\dots&\dots& \Pi^{\text{N\'eel}}_0
  \end{pmatrix}, 
\end{align}
where we have assumed the subsystem size $\ell$ to be even and
\be
\Pi^{\text{N\'eel}}_l = \begin{pmatrix} \avg{a_1 a_{4l+1}}-\delta_{0l} & \avg{a_2 a_{4l+1}} & \avg{a_3 a_{4l+1}} & \avg{a_0 a_{4l-3}} \\
\avg{a_1 a_{4l+2}} & \avg{a_2 a_{4l+2}}-\delta_{l0} &\avg{a_3 a_{4l+2}} & \avg{a_0 a_{4l-2}} \\
\avg{a_1 a_{4l+3}} & \avg{a_2 a_{4l+3}} & \avg{a_3 a_{4l+3}}-\delta_{l0} & \avg{a_0 a_{4l-1}} \\
\avg{a_{1} a_{4l+4}} & \avg{a_2 a_{4l+4}} & \avg{a_3 a_{4l+4}} &
\avg{a_0 a_{4l}}-\delta_{l0} 
\end{pmatrix}
=\begin{pmatrix} 
-f_l & g_l & h_l & 0 \\
-g_{-l} & f_l & 0 & h_{l} \\
-h_{-l} & 0 &f_l & -g_l \\
0 & -h_{-l}& g_{-l} & -f_l
\end{pmatrix} -\delta_{l,0}\mathbbm{1}\ .
\ee
Here the various two point functions are given by
\bea
f_l -\delta_{l0} &=&i \int_{0}^{2\pi} \frac{\id{k}}{2\pi} e^{-2ijk}
\Big[ e^{i\theta_k} \cos\big(\eps(k+\pi)t\big)\sin\big(\eps(k)t\big)-
e^{-i\theta_{k+\pi}} \cos\big(\eps(k)t\big)\sin\big(\eps(k+\pi)t\big)\Big] \ ,\nn
g_l &=& i \int_0^{2\pi} \frac{\id{k}}{2\pi} e^{-2ijk}
\Big[\cos\big(\eps(k)t\big)\cos\big(\eps(k+\pi)t\big) +
e^{i(\theta_k+\theta_{k+\pi})} \sin\big(\eps(k)t\big)\sin\big(\eps(k+\pi)t\big)
\Big],\nn
h_l &=& i \int_0^{2\pi} \frac{\id{k}}{2\pi} e^{-i(2j-1)k}
\Big[e^{-i\theta_k}\cos\big(\eps(k+\pi)t\big)\sin\big(\eps(k)t\big)
-e^{i\theta_{k+\pi}}\cos\big(\eps(k)t\big)\sin\big(\eps(k+\pi)t\big)\Big].
\eea
In the following we will determine the characteristic functions
\be
\chi^{(u)}(\lambda,\ell,t)=\langle\psi_0(t)|e^{i\lambda
  S^z_u(\ell)}|\psi_0(t)\rangle\ ,\qquad
\chi^{(s)}(\lambda,\ell,t)=\langle\psi_0(t)|e^{i\lambda
  S^z_s(\ell)}|\psi_0(t)\rangle \ ,
\ee
where again we have defined
\be
S^z_u(\ell)=\sum_{j=1}^\ell \sigma^z_j\ ,\quad
S^z_s(\ell)=\sum_{j=1}^\ell (-1)^j\sigma^z_j\ .
\ee
According to our general discussion in section \ref{Sec2} they have determinant representations of the form
\bea
\chi^{(u)}(\lambda,\ell,t)&=&\big(2\cos(\lambda)\big)^\ell
\sqrt{{\rm
    det}\Big(\frac{1+\Gamma^A_{\text{N\'eel}}\tilde{\Gamma}}{2}\Big)}\ ,\nn
\chi^{(s)}(\lambda,\ell,t)&=&\big(2\cos(\lambda)\big)^\ell
\sqrt{{\rm
    det}\Big(\frac{1+\Gamma^A_{\text{N\'eel}}\tilde{\Gamma}^{\pi}}{2}\Big)}\ ,
\label{detrep}
\eea
where
$\tilde\Gamma^\pi_{2j,2j-1}=-\tilde\Gamma^\pi_{2j-1,2j}=-\tan(\lambda)(-1)^j$ and
$\tilde\Gamma_{2j,2j-1}=-\tilde\Gamma^\pi_{2j-1,2j}=-\tan(\lambda)$ respectively.

%%%%%%%%%%%%%%%%%%%%%%%%%%%%%%%%%%%%%%%%%%%%%%%%%%%
\subsubsection{Behaviour in the stationary state}
%%%%%%%%%%%%%%%%%%%%%%%%%%%%%%%%%%%%%%%%%%%%%%%%%%%
We first consider the probability distributions for a finite subsystem
of even size $\ell$ in the late time limit. As we will now show, the
stationary state for the N\'eel quench is locally equivalent to an
infinite temperature state. To see this we first note that the energy
of the N\'eel state is 
\be
\langle\psi_0|H(h)|\psi_0\rangle=0.
\ee
It is easy to see using their explicit representation in terms of
spins \cite{EFreview} that the expectation values of all higher
conservation laws also vanish
\be
\langle\psi_0|I^{(n,\pm)}|\psi_0\rangle=0.
\ee
This in turn implies that the conserved Bogoliubov mode occupation numbers %, \emph{cf.} Appendix~\ref{Appendix:TFIC}, 
are given by
\be
\langle\psi_0|\alpha^\dagger_k\alpha_k|\psi_0\rangle=\frac{1}{2}.
\ee
These characterize an infinite temperature equilibrium state. We conclude
that the system will relax locally \cite{EFreview} to an infinite
temperature steady state at late times after the quench. Using this
observation it is then straightforward to work out the probability
distributions $P^{(u)}(m,t=\infty)$ of $S^z_u(\ell)=\sum_{j=1}^\ell \sigma^z_j$ and
$P^{(s)}(m,\infty)$ of $S^z_s(\ell)=\sum_{j=1}^\ell (-1)^j\sigma^z_j$. As shown
in the introduction we have
\be
P^{(u,s)}(m,t=\infty)=2\sum_{{r\in\mathbb{Z}}}P^{(u,s)}_w(r) 
\begin{cases} 
\delta(m-2r+\ell) & \ell \text{ odd} \\  
\delta(m-2r)  & \ell \text{ even}
\end{cases}\ .
\ee
As we are dealing with an infinite temperature state, we may calculate 
$P_w(r)$ by using a grand canonical ensemble and working in the
simultaneous eigenbasis of the $\sigma^z_j$'s. This reduces the
calculation of $P_w(r)$ to the combinatorial problem of how many
eigenstates there are for a given eigenvalue of $S^z_u(\ell)$ or
$S^z_s(\ell)$. This is easily solved in terms of the binomial distribution
\begin{align}
P^{(u)}_w(m,\infty) =P_w^{(s)}(m,\infty) 
=\frac{1}{2^\ell}\binom{\ell}{\ell/2-m} 
\sim \sqrt{\frac{2}{\ell\pi}} \exp{-\frac{2m^2}{\ell}} \ .
\label{eq:Neel_asymptotic_prob}
\end{align}
The result \fr{eq:Neel_asymptotic_prob} for large $\ell$ is of course
reproduced by applying Szeg\H{o}'s Lemma for block Toeplitz matrices
to the determinant representations \fr{detrep}. This gives
\be
\lim_{t\to\infty} \frac{\ln\chi^{(u)}(\lambda,\ell,t)}{\ell}= 
\frac{1}{2}\ln\big(\cos^2(\lambda)\big)+{\cal O}(1/\ell)
=\lim_{t\to\infty} \frac{\ln\chi^{(s)}(\lambda,\ell,t)}{\ell}
\ ,\quad \ell\gg 1.
\ee
Fourier transforming gives the Gaussian form of $P^{(u,s)}_w(m,\infty)$ in
\fr{eq:Neel_asymptotic_prob}.
%%%%%%%%%%%%%%%%%%%%%%%%%%%%%%%%%%%%%%%%%%%%%%%%%%%
\subsubsection{Time dependence}
\begin{figure}[ht!]
(a)\includegraphics[width=0.45\textwidth]{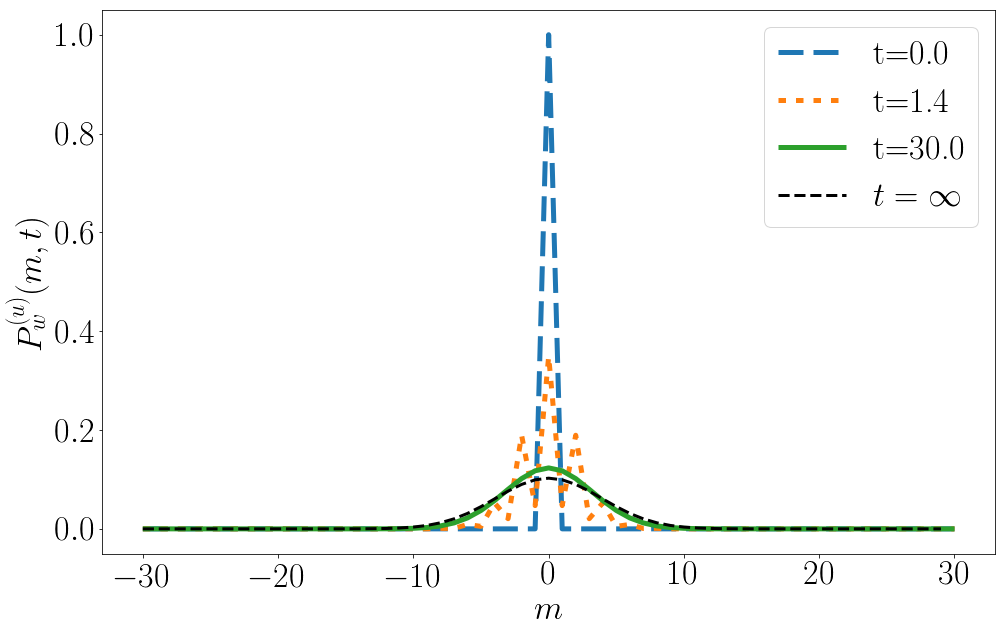}\qquad
(b)\includegraphics[width=0.45\textwidth]{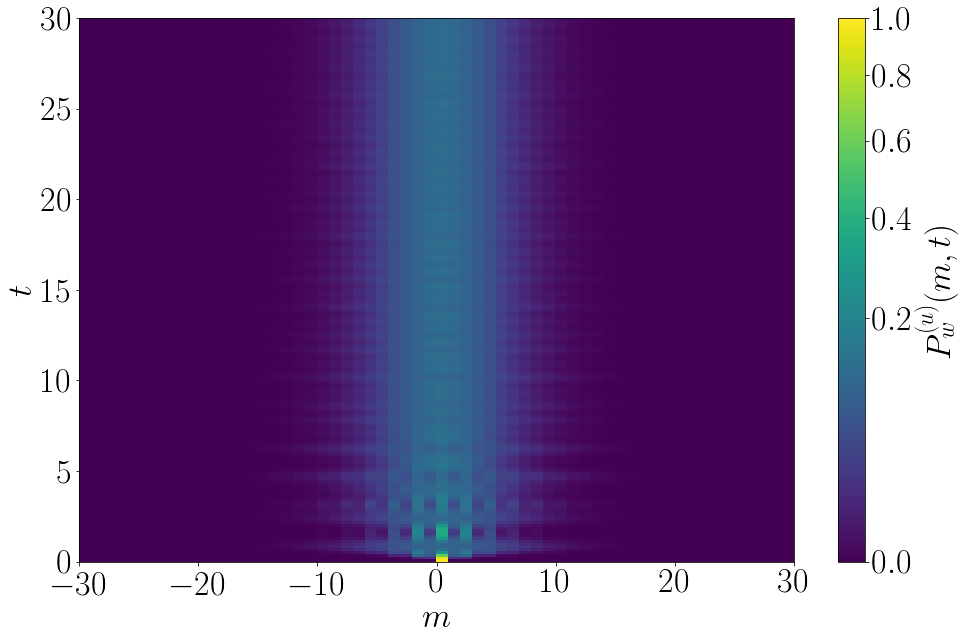}
\caption{$P^{(u)}_w(m,t)$ for a subsystem of size $\ell=60$
at times $t=0,1.4,30.0$ for a system initialized in a N\'eel
state and time evolved with $H(h=0.2)$. The dotted lines are the
asymptotic probability distributions given in
\eqref{eq:Neel_asymptotic_prob}.}
\label{fig:NeelQ1}
\end{figure}

\begin{figure}[ht!]
(a)\includegraphics[width=0.45\textwidth]{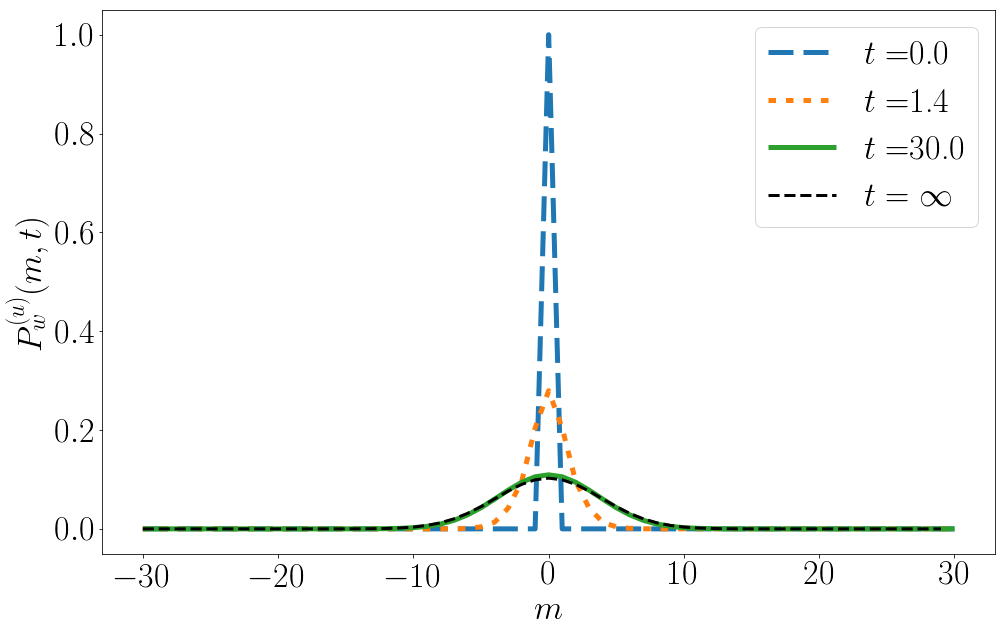}
(b)\includegraphics[width=0.45\textwidth]{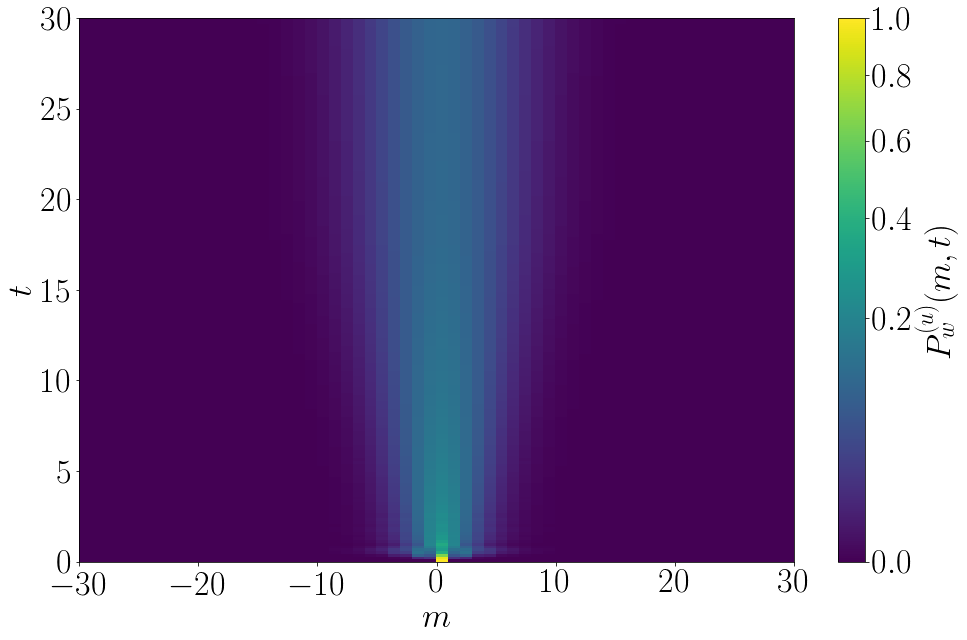}
\caption{$P^{(u)}_w(m,t)$ for a subsystem of size $\ell=60$
at times $t=0,1.4,30.0$ for a system initialized in a N\'eel
state and time evolved with $H(h=2)$. The dotted lines are the
asymptotic probability distributions given in
\eqref{eq:Neel_asymptotic_prob}.}
\label{fig:NeelQ2}
\end{figure}

\begin{figure}[ht!]
(a)\includegraphics[width=0.45\textwidth]{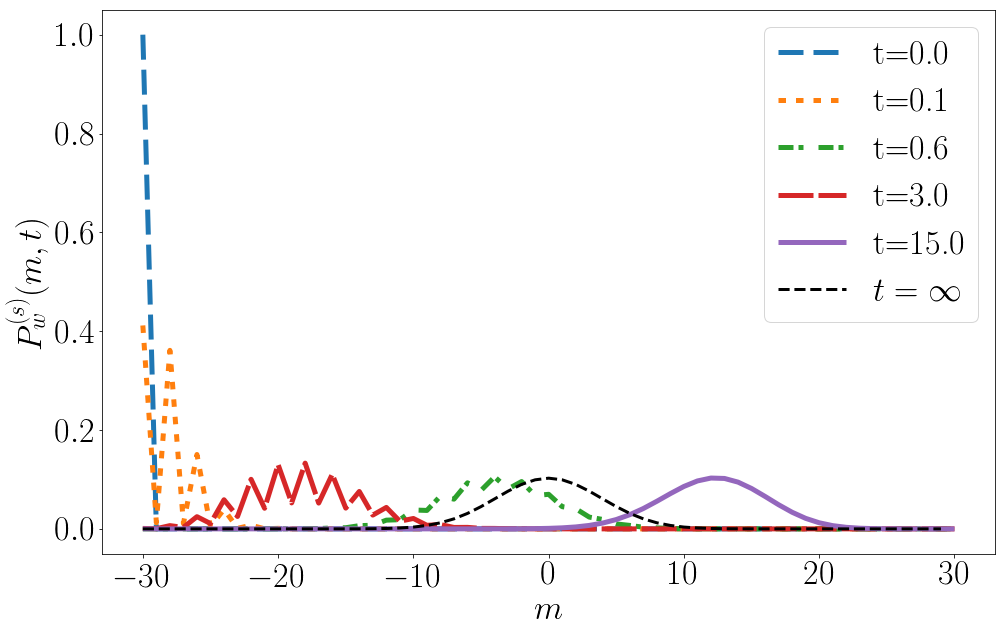}
(b)\includegraphics[width=0.45\textwidth]{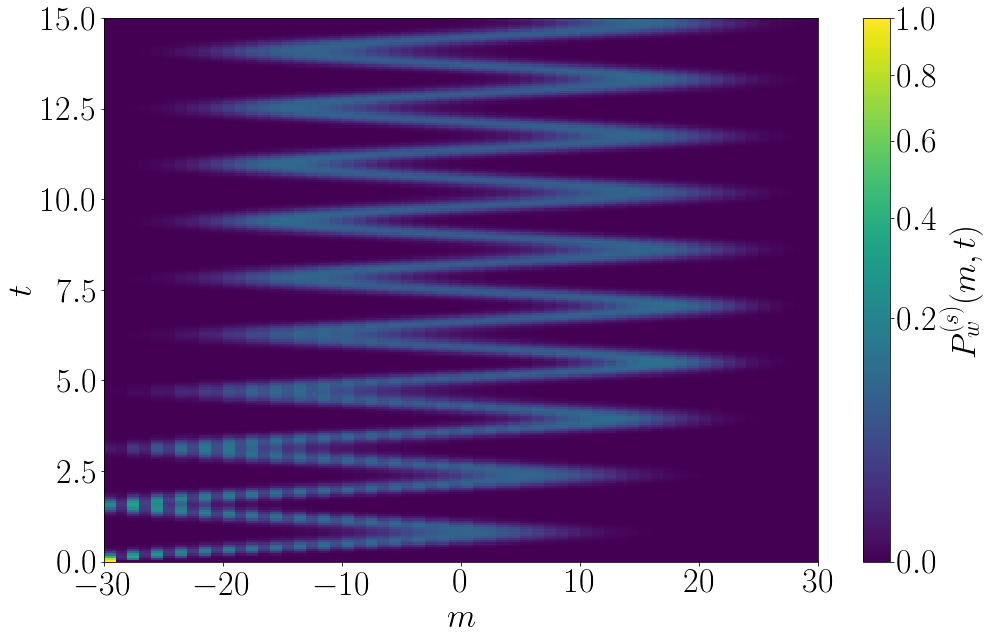}
\caption{$P^{(s)}_w(m,t)$ for a subsystem of size $\ell=60$
at times $t=0,1.4,30.0$ for a system initialized in a N\'eel
state and time evolved with $H(h=0.2)$. The dotted lines are the
asymptotic probability distributions given in
\eqref{eq:Neel_asymptotic_prob}.}
\label{fig:NeelQ3}
\end{figure}

\begin{figure}[ht!]
(a)\includegraphics[width=0.45\textwidth]{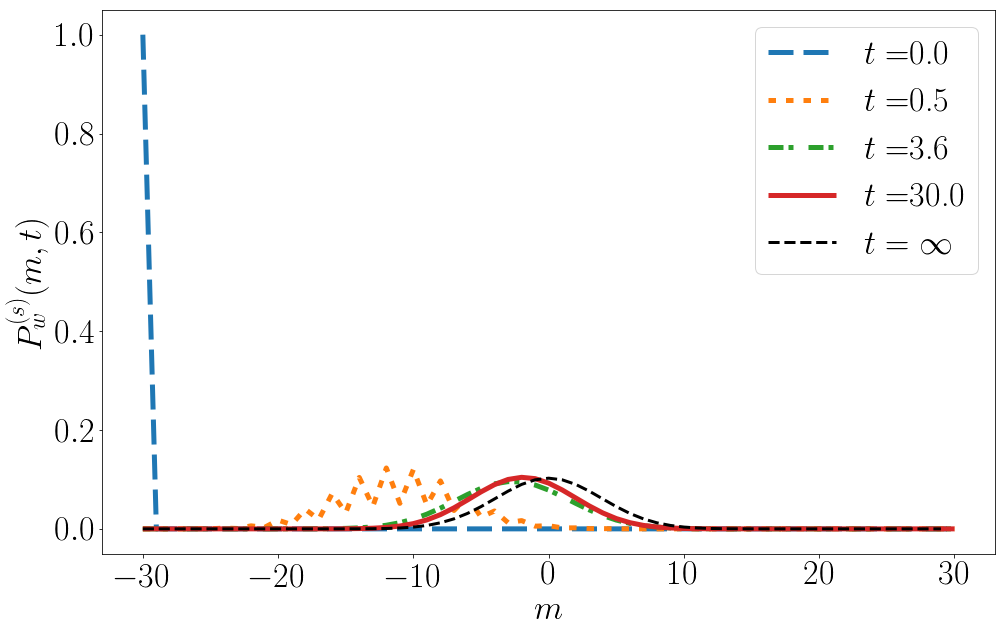}
(b)\includegraphics[width=0.45\textwidth]{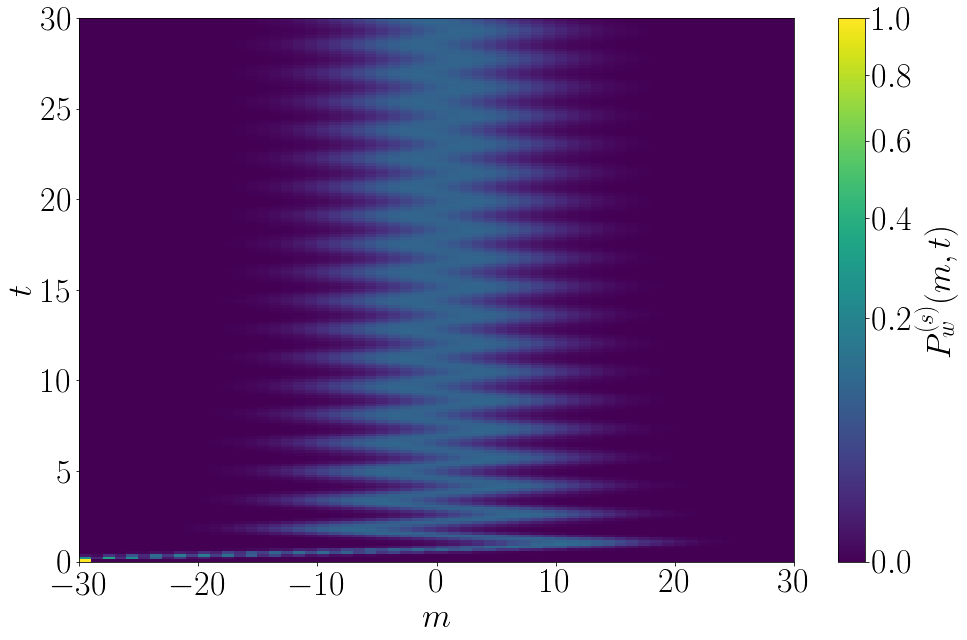}
\caption{$P^{(s)}_w(m,t)$ for a subsystem of size $\ell=60$
at times $t=0,1.4,30.0$ for a system initialized in a N\'eel
state and time evolved with $H(h=2)$. The dotted lines are the
asymptotic probability distributions given in
\eqref{eq:Neel_asymptotic_prob}.}
\label{fig:NeelQ4}
\end{figure}

%%%%%%%%%%%%%%%%%%%%%%%%%%%%%%%%%%%%%%%%%%%%%%%%%%%
The time dependence of the probability distributions for both
$S^z_u(\ell)$ and $S^z_s(\ell)$ can now be determined numerically from
the determinant representation \fr{detrep}. Results for two values of
the transverse field ($h=0.2$ and $h=2$) are shown in
Figs~\ref{fig:NeelQ1}, \ref{fig:NeelQ2}, \ref{fig:NeelQ3} and
\ref{fig:NeelQ4}. 
The probability distribution of $S^z_u(\ell)$ initially has a single
peak at $m=0$. At later times this peak broadens and relaxes towards
the Gaussian profile \fr{eq:Neel_asymptotic_prob}. When quenching to
the ferromagnetic phase, \emph{cf.} Fig.~\ref{fig:NeelQ1}, an
additional feature emerges: an even/odd structure evolves at short
times after the quench.

The probability distribution of $S^z_s(\ell)$ is useful for investigating
the restoration of the translational symmetry. In the initial state
$P_w^{(s)}(m,t=0)$ features a single peak at $m=-\ell/2$, which
is a characteristic fingerprint of the classical N\'eel state (in
z-direction). We first discuss quenches into the ferromagnetic phase. 
Here at short times after the quench $P_w^{(s)}(m,t)$
develops an even/odd structure and broadens significantly. The average
of the probability distribution oscillates strongly in time and decays
very slowly to its stationary value, which is a Gaussian distribution
centred around $m=0$. This shows that translational symmetry is
restored very slowly. 

The behaviour for quenches into the paramagnetic phase is broadly
similar. An even/odd structure develops at early times, but is less
pronounced that for quenches to the ferromagnetic phase. The average
of $P_w^{(s)}(m,t)$ again oscillates strongly around $m=0$, but is
seen to relax much more quickly than for quenches to the ferromagnetic
phase. Approximate translational symmetry gets restored more rapidly.

%%%%%%%%%%%%%%%%%%%%%%%%%%%%%%%%%%%%%%%%%%%%%%%%%%%%%%%%%%%%%%%%%%%%%%%%%%
%%%%%%%%%%%%%%%%%%%%%%%%%%%%%%%%%%%%%%%%%%%%%%%%%%%%%%%%%%%%%%%%%%%%%%%%%%
\section{Analytic results for the probability distribution}
\label{sec:analytic}
%%%%%%%%%%%%%%%%%%%%%%%%%%%%%%%%%%%%%%%%%%%%%%%%%%%%%%%%%%%%%%%%%%%%%%%%%%
%%%%%%%%%%%%%%%%%%%%%%%%%%%%%%%%%%%%%%%%%%%%%%%%%%%%%%%%%%%%%%%%%%%%%%%%%%
We now restrict our discussion to the particular case of transverse
field quenches. As we have seen above, in this case the characteristic
functions $\chi^{(u)}(\lambda,\ell,t)$ exhibit a scaling collapse at late
times, \emph{cf.} \fr{scaling}. This suggests that it might be
possible to obtain analytic results for the late time asymptotics by a
suitable generalization of the multi-dimensional stationary
state approximation method previously used to determine the
asymptotics of the order parameter two-point function \cite{CEF1} and
the entanglement entropy \cite{FC08}. As we will see, such a
generalization is indeed possible, even though the case at hand is
significantly more complicated.

Our starting point is the following expression
\be
\ln\chi^{(u)}(\lambda,\ell,t) = \ell \ln\left(\cos\lambda\right) +\frac{1}{2}
  \tr{\ln(1-\tan\lambda \; \Gamma^\prime)}\ ,
\label{starting}
\ee
which is derived from \eqref{chiTI} by using the identity
$\ln\left(\det{A}\right)=\tr{\ln\left(A\right)}$.
The second term in \fr{starting} can be expanded in a power series
\begin{align}
\frac{1}{2}\tr{\ln(1-\tan\lambda \; \Gamma^\prime)} =& -\frac{1}{2}
\sum_{n=1}^\infty \frac{\big(\tan(\lambda)\big)^n}{n}
\ {\rm Tr}\big[(\Gamma^{\prime})^n\big] \ . \label{chipowerseries}
\end{align}
This then leads us to examine integer powers $(\Gamma^{\prime})^n$ of
the correlation matrix. Unlike in the case of the order parameter
two-point function analyzed in \cite{CEF2} odd powers do not vanish 
because $\Gamma^\prime$ is not a real anti-symmetric matrix. The
\emph{symbol} $t^\prime(k)$ corresponding to the correlation matrix
$\Gamma'$ is defined by
\be
\left(\Gamma^\prime\right)_{ln} = 
\int_{-\pi}^\pi \frac{\id{k}}{2\pi} e^{i(l-n)k} \;\hat
t^\prime(k)\ .
\ee
Its explicit expression for a magnetic field quench from $h_0$ to $h$ is
\be
\hat t^\prime(k) = 
\begin{pmatrix} -i e^{i\theta_k} (\cos\Delta_k - i \sin\Delta_k
    \cos(2\eps_k t)) & \sin\Delta_k\sin(2\eps_kt) \\
    \sin\Delta_k \sin(2\eps_k t) & -i e^{-i\theta_k}(\cos\Delta_k +i\sin\Delta_k
    \cos(2\eps_k t))
\end{pmatrix} \ ,
\ee
%\\ 
%  =& \sin\Delta_k \sin(2\eps_k t) \sigma_x
%  -i(\cos\Delta_k\cos\theta_k+\cos(2\eps_k
%  t)\sin\Delta_k\sin\theta_k)\mathbbm{1} 
%  \notag \\ & \hspace{2cm}+
%  (\cos\Delta_k\sin\theta_k-\cos(2\eps_kt)\cos\theta_k\sin\Delta_k) \sigma_z
%\end{align}
where $\theta_k$ and $\Delta_k$ have been previous defined in \fr{thetaDelta}.
Following Ref.~\cite{CEF2} we can represent the trace of powers of the
correlation matrix as multiple integrals
\be
{\rm Tr}\big[(\Gamma^\prime)^n\big] = 
\left(\frac{\ell}{2}\right)^{n} \int_{-\pi}^{\pi}
\frac{dk_1\dots dk_n}{(2\pi)^n}\int_{-1}^1d\xi_1\dots d\xi_n\ 
C\big(\vec{k}\big)\ F\big(\vec{k}\big)\
{\rm exp}\left(i\ell \sum_{j=0}^{n-1} \frac{\xi_j}{2}(k_{j+1}-k_j)\right),
\ee
where we have defined $k_0\equiv k_n$ and
\be
C(\vec{k})=\prod_{j=0}^{n-1}
\frac{k_j-k_{j-1}}{2\sin\left[(k_j-k_{j-1})/2\right]}\ ,\qquad
F(\vec{k})=\tr{\prod_{i=0}^{n-1}\hat t^\prime(k_i)}\ .
\ee
We now change variables
\be
\zeta_0 = \xi_1\ ,\quad
\zeta_i = \xi_{i+1}-\xi_i\ ,\ i=1,\dots,n-1\ .
\ee
The integration ranges in the $\zeta$ variables is determined by the
constraints
\be
-1\leq\sum_{j=0}^{k-1}\zeta_j\leq 1\ ,\quad k=1,\dots,n.
\label{constraints}
\ee
The integral over $\zeta_0$ can now be carried out as the integrand
does not depend on it. This gives
\be
{\rm Tr}\big[(\Gamma^\prime)^n\big] = 
\left(\frac{\ell}{2}\right)^{n} \int_{-\pi}^{\pi}
\frac{dk_1\dots dk_n}{(2\pi)^n}\int_{-1}^1d\zeta_1\dots d\zeta_{n-1}\ 
\mu(\vec\zeta)\ C\big(\vec{k}\big)\ F\big(\vec{k}\big)\
{\rm exp}\left(-i\ell \sum_{j=1}^{n-1} \frac{\zeta_j}{2}(k_j-k_0)\right),
\label{zeta}
\ee
where $\mu(\{\zeta\})$ is the size of the range of $\zeta_0$ under the
constraints \fr{constraints}
\be
\mu(\vec{\zeta}) = \max\left[0,\min_{0\leq j\leq n-1}\left(1-\sum_{k=1}^j
\zeta_k\right)+\min_{0\leq j\leq n-1}\left(1+\sum_{k=1}^j\zeta_k\right)\right].
\ee
%%%%%%%%%%%%%%%%%%%%%%%%%%%%%%%%%%%%%%%%%%%%%%%%%%%%%%%%%%%%%
\subsection{Multi-dimensional stationary phase approximation}
%%%%%%%%%%%%%%%%%%%%%%%%%%%%%%%%%%%%%%%%%%%%%%%%%%%%%%%%%%%%%
For large values of $\ell$ the integrals can be carried out using a
multi-dimensional stationary phase approximation. As the symbol is
independent of $\zeta_j$ the stationarity conditions for the
$\zeta_j$'s implies that the leading contribution to \fr{zeta} derives 
from the region
\be
k_j\approx k_0\ ,\quad j=1,\dots,n-1\ .
\ee
We may thus replace $k_j$ with $k_0$ everywhere except in rapidly
oscillating terms in the symbol such as $e^{2i\eps(k_j)t}$. In
\cite{CEF2} this procedure was referred to as \textit{localisation 
  rule}. As in \cite{CEF1} application of this rule gives 
\be
C(\vec{k})\approx 1\ .
\ee
Obtaining a closed form expression for $F(\vec{k})$ is however much more
involved than for the order-parameter two point function studied in 
Ref.~\cite{CEF2}. We conjecture that application of the localization rule to
$F(\vec{k})$ results in
\begin{align}
F(\vec{k})\Big|_{\rm loc} =& 2\sum_{A_1,A_2,A_3 }
\text{sign}(A_1\cup A_2) (-i)^{n+S(A_1,A_2)}\
(\cos\Delta_{k_0})^{|A_3|}
(\sin\Delta_{k_0})^{|A_1|+|A_2|} \notag \\ &\times
\cos\Big([n-2q(A_1)]\theta_{k_0}-\frac{\pi(|A_1|+|A_2|)}{2}\Big)
\prod_{i\in A_1} \sin(2\eps(k_i)t) \prod_{j\in A_2} \cos(2\eps(k_j)t)
\ .
\label{Floc}
\end{align}
Here the sum is over all partitions of the set of integers
$\{0,1,\dots,n-1\}$ into three sets $A_{1}$, $A_2$ and $A_3$, where the
number of elements in $A_1$ is constrained to be even. The size of the
set $B=\{b_1,b_2,\dots\}$ is denoted by $|B|$ and we have defined
\bea
q(B)&=&\text{mod}_n\big[\sum_{i=1}^{\abs{B}}(-1)^{i+1}b_{i}\big] ,\nn
S(A_1,A_2)&=& \begin{cases} 
2 & \text{if } q(A_1)\leq \frac{n}{2},\ \text{mod}_2\big[
\abs{A_1\cup A_2}\big]=1
\text{ and } \abs{A_1}>0,\\ 
0 & \text{else}.
\end{cases}
\eea
Finally, $\text{sign}(A)$ is the sign of the permutation required to
bring the (integer) elements of the set $A$ into ascending order.
We have explicitly checked \fr{Floc} for $1\leq n\leq 15$ but have not
been able to find a rigorous proof for it.

We now use the identity (for even $k$)
\begin{align}
\prod_{i=1}^k \sin(x_i)\prod_{j=k+1}^{k+m}\cos(x_j) =& 
\frac{(-1)^{\frac{k}{2}}}{2^{k+m}} 
\sum_{i_1=0}^1\sum_{i_2=0}^1\dots\sum_{i_{k+m}=0}^1
{\rm exp}\big(i\sum_{j=1}^{k+m}(-1)^{i_j}x_j+i\pi\sum_{j=1}^ki_j\big)\ ,
\end{align}
to rewrite the time-dependent factors in \fr{Floc}. This gives 
\begin{align}
F(\vec{k})\Big|_{\rm loc} =&
2 \sum_{A_1,A_2,A_3 }\text{sign}(A_1\cup A_2)\
\frac{(-i)^{S(A_1,A_2)+n+|A_1|}}{2^{|A_1|+|A_2|}}
(\cos\Delta_{k_0})^{|A_3|} (\sin\Delta_{k_0})^{|A_1|+|A_2|} \notag
\\ &\times 
\cos\Big([n-2q(A_1)]\theta_{k_0}-\frac{\pi(|A_1|+|A_2|)}{2}\Big)
\notag \\ & \times \sum_{p_1=0}^1\dots\sum_{p_{|A_1|+|A_2|}=0}^1
{\rm exp}\Big[2it\sum_{r=1}^{|A_1|+|A_2|}(-1)^{p_r}\eps(k_{(A_1\cup
    A_2)_r})+i\pi\sum_{r=1}^{\abs{A_1}}p_{r}\Big]\ ,
\end{align}
where $(A)_r$ is the r'th element of the set $A$ and
\be
(A_1\cup A_2)_r=
\begin{cases}
(A_1)_r & \text{if } r\leq|A_1|,\\
(A_2)_{r-|A_1|} & \text{if } |A_1|<r\leq|A_1|+|A_2|.
\end{cases}
\ee
Application of the localization rule to \fr{zeta} hence results in an
expression of the form
\bea
{\rm Tr}\big[(\Gamma^\prime)^n\big]\Big|_{\rm loc} &=& 
2\left(\frac{\ell}{2}\right)^{n} 
 \sum_{A_1,A_2,A_3 }\text{sign}(A_1\cup A_2)\
\frac{(-i)^{S(A_1,A_2)+n+|A_1|}}{2^{|A_1|+|A_2|}}
(\cos\Delta_{k_0})^{|A_3|} (\sin\Delta_{k_0})^{|A_1|+|A_2|}\nn
&&\times
\cos\Big([n-2q(A_1)]\theta_{k_0}-\frac{\pi(|A_1|+|A_2|)}{2}\Big)
\sum_{p_1=0}^1\dots\sum_{p_{|A_1|+|A_2|}=0}^1(-1)^{\sum_{r=1}^{\abs{A_1}}p_{r}}
\int_{-1}^1d\zeta_1\dots d\zeta_{n-1}\ \mu(\vec\zeta)\nn
&&\times\int_{-\pi}^{\pi}
\frac{dk_1\dots dk_n}{(2\pi)^n}\
{\rm exp}\Big[2it\sum_{r=1}^{|A_1|+|A_2|}(-1)^{p_r}\eps(k_{(A_1\cup
    A_2)_r})-i\ell \sum_{j=1}^{n-1}\frac{\zeta_j}{2}(k_j-k_0)\Big]\ .
\label{zetaloc}
\eea
In the next step we carry out a multi-dimensional stationary phase
approximation for the $2n-2$ integrals over $\zeta_1,\dots,\zeta_{n-1}$ and
$k_1,\dots,k_{n-1}$. We will assume that there is a single saddle
point and use 
\be
\int dx_1\dots dx_k\ p(x_1,\dots,x_k) 
e^{i\ell q(x_1,\dots,x_k)} \approx
\left( \frac{2\pi}{\ell} \right)^{k/2}
\frac{p(x^{(0)}_1,\dots,x^{(0)}_k)}{\sqrt{|\det{A}|}}
\exp{i\ell q(x^{(0)}_1,\dots,x^{(0)}_k)+\frac{i\pi\sigma_A}{4}},
\ee
where $\sigma_A$ the signature of the matrix $A$ (i.e. the difference
between the numbers of positive and negative eigenvalues), which is
the Hessian of the function $q$ evaluated at the saddle point
\be
A_{ij}=\frac{\partial}{\partial x_i}
\frac{\partial}{\partial
  x_j}\Bigg|_{\vec{x}=\vec{x}^{(0)}}q(x_1,\dots,x_k)\ .
\ee 
In our case the saddle point conditions are
\bea
k^{(0)}_j &=& k_0\ ,\quad j=1,\dots,n-1\ ,\nn
\zeta^{(0)}_j &=& \begin{cases}
\gamma_{A_1\cup A_2,k}& \text{if } j\in A_1\cup A_2\ , \\
0 & \text{else}\ ,
\end{cases}
\eea
where 
$\gamma_{A,k}=\frac{4t}{\ell}(-1)^{p_{(A)^{-1}_k}}\eps^\prime(k_0)$ and 
$(A_1\cup A_2)^{-1}$ is the inverse of the index-function
$(A_1\cup A_2)_j$ defined above. The Hessian $A$ is a matrix of the form  
\begin{align}
A=& \frac{1}{2}\begin{pmatrix} 0 & I \\ I & M 
\end{pmatrix}\ ,
\end{align}
and hence we have $\det{A}=-4^{1-n}$ and $\sigma_A=0$. The value of
$\mu(\vec{\zeta})$ at the saddle point for a given sequence
$\{p_1,p_2,\dots,P_{|A_1|+|A_2|}\}$ is 
\be
\mu(\vec{\zeta}^{(0)})=
\max\left[0, \min_{0\leq j\leq |B|}\left( 1-\sum_{k=1}^j \gamma_{B,k}
\right)+ \min_{0\leq j\leq |B|}\left( 1+\sum_{k=1}^j 
\gamma_{B,k}\right)\right],
\ee
where $B=A_1\cup A_2-\{0\}$. The saddle point approximation thus gives
\bea
{\rm Tr}\big[(\Gamma^\prime)^n\big] &\approx& 
\ell \sum_{A_1,A_2,A_3 }\text{sign}(A_1\cup A_2)\
\frac{(-i)^{S(A_1,A_2)+n+|A_1|}}{2^{|A_1|+|A_2|}}
(\cos\Delta_{k_0})^{|A_3|} (\sin\Delta_{k_0})^{|A_1|+|A_2|}\nn
&&\times
\cos\Big([n-2q(A_1)]\theta_{k_0}-\frac{\pi(|A_1|+|A_2|)}{2}\Big)
\sum_{p_1=0}^1\dots\sum_{p_{|A_1|+|A_2|}=0}^1(-1)^{\sum_{r=1}^{\abs{A_1}}p_{r}}\nn
&&\times\int_{-\pi}^\pi\frac{dk_0}{2\pi}\
 \mu(\vec\zeta^{(0)})\ 
{\rm exp}\Big[-2it\eps(k_0)\sum_{r=1}^{\abs{A_1\cup A_2}}(-1)^{p_r}\Big].
\label{SP}
\eea
The leading contribution to the final integral can then also be
determined by a stationary phase approximation. This shows that all
terms with $\sum_{r=1}^{\abs{A_1\cup A_2}}(-1)^{p_r}\neq0$ are
suppressed at late times by a factor of $1/\sqrt{t}$. Conversely, the
leading contribution to $\chi^{(u)}(\lambda,\ell,t)$ at late times arises
from terms with $\sum_{r=1}^{\abs{A_1\cup A_2}}(-1)^{p_r}=0$, which
requires $|A_1|+|A_2|$ to be even. 
%%%%%%%%%%%%%%%%%%%%%%%%%%%%%%%%%%%%
\subsubsection{Structure of   $\mu(\vec{\zeta}^{(0)})$}\label{structureofmu}
%%%%%%%%%%%%%%%%%%%%%%%%%%%%%%%%%%%%
At this point it is useful to investigate the structure of
$\mu(\vec{\zeta}^{(0)})$ for a given term in the multiple sum over
$p_1,\dots,p_{|A_1|+|A_2|}$ in more detail. For simplicity we focus on
a particular example
\be
|A_1|=|A_2|=2\ ,\quad
\{p_{(A_1\cup A_2)^{-1}_k}|k=1,\dots,4\}=\{0,1,0,1\}.
\ee
In this case we have
\begin{align}
  \mu(\vec{\zeta}^{(0)})=& \max\Big(0,\min(1,1-\frac{4t}{\ell}\eps^\prime(k_0))+\min(1,1+\frac{4t}{\ell}\eps^\prime(k_0))\Big) \notag \\
  =& \max\Big(0,2-\frac{4t}{\ell}\abs{\eps^\prime(k_0)}\Big)=
\Theta(\ell-2\abs{v_{k_0}}t) \Big(2-4 \frac{t\abs{v_{k_0}}}{\ell}\Big),
\label{LS2}
\end{align}
where $v_{k_0}=\eps^\prime(k_0)$ is the group velocity of Bogoliubov
fermions at momentum $k_0$ and $\Theta(x)$ is the Heaviside step function. The
step function in \fr{LS2} is reminiscent of the light-cone structure found for two point correlation functions of local 
operators \cite{cc-06,cc-07,bel-14,BEGR-16} and entanglement entropies \cite{cc-05,ac-17,ac-18}. 
Repeating the above exercise for
\be
|A_1|=|A_2|=m\ ,\quad
\{p_{(A_1\cup A_2)^{-1}_k}|k=1,\dots,2m\}=\{\underbrace{1,1,\dots,1}_{m},0,0,\dots,0\}\ ,
\ee
leads to the result
\be
\mu(\vec{\zeta}^{(0)}) = \Theta(\ell-2m\abs{v_{k_0}}t) \Big(2-4m
\frac{t\abs{v_{k_0}}}{\ell}\Big)\ .
\ee
All other cases can be worked out analogously and lead to 
Heaviside step functions $\Theta(\ell-2m\abs{v_{k_0}}t)$ with
$m\in\mathbb{N}_0$. 

%%%%%%%%%%%%%%%%%%%%%%%%%%%%%%%%%%%%
\subsection{Result for $\chi(\lambda,\ell,t)$}
%%%%%%%%%%%%%%%%%%%%%%%%%%%%%%%%%%%%
In order to obtain the logarithm of the characteristic function
$\chi(\lambda,\ell,t)$ we now need to sum over all contributions
\fr{SP} with coefficients given in \fr{chipowerseries}. This is a
formidable task. It turns out that the structure of Heaviside step
functions discussed above provides a very useful way of organizing the
complicated summation required. The full result can be expressed in
the form
\be
\ln{\chi^{(u)}(\lambda,\ell,t)}\approx \ell \ln(\cos\lambda) + \frac{\ell}{2} 
\sum_{n=0}^\infty \int_0^{2\pi}\frac{dk_0}{2\pi} \Theta(\ell-2n|v_k|t)
\left[1-\frac{2n|v_k|t}{\ell}\right]
\sum_{m=0}^{n+1}
\cos\big(2m\eps(k_0)t\big)f_{n,m}(\lambda,k_0)+{\cal C}\ .
\label{chiSP}
\ee
Here ${\cal C}$ is a constant that is beyond the accuracy of the
stationary phase approximation and the functions
$f_{n,m}(\lambda,k_0,t)$ are given in terms of infinite series. Based
on the first 15 terms in these series we conjecture the following
explicit expressions
\bea
f_{0,0}(\lambda,k_0)&=&2 \ln\big(1+i\cos\Delta_{k_0}
\tan{\lambda}e^{i\theta_{k_0}} \big)\ ,\nn
f_{1,0}(\lambda,k_0)&=&\ln\left[ 
1-\frac{\sin^2\Delta_{k_0}\tan^2\lambda(\cos\theta_{k_0}+i\cos\Delta_{k_0}
\tan\lambda)^2}{(\sin^2\theta_{k_0}+(\cos\theta_{k_0}+i\cos\Delta_{k_0}
\tan\lambda)^2)^2} \right]\ ,\nn
f_{2,0}(\lambda,k_0)&=&\ln\Bigg[1+\frac{\sin^4\Delta_{k_0}
\tan^4\lambda\sin^2\theta_{k_0}(\cos\theta_{k_0}+i\cos\Delta_{k_0}
\tan{\lambda})^2}{((\sin^2\theta_{k_0}+(\cos\theta_{k_0}
+i\cos\Delta_{k_0}\tan\lambda)^2)^2-\sin^2\Delta_{k_0}
\tan^2\lambda(\cos\theta_{k_0}+i\cos\Delta_{k_0}\tan\lambda)^2)^2}\Bigg].\nn
\label{fn0}
\eea
In principle one could determine further terms $f_{n,0}$ but their
contribution turns out to be negligible for all cases we have
considered. The contributions $f_{n,m>0}(\lambda,k_0,t)$ are more
difficult to simplify. While the term $f_{0,1}$ can still be obtained without further approximations, in order to obtain closed form expressions for $m>1$ we
have resorted to an expansion in powers of $\sin(\Delta_{k_0})$. This
is expected to give very accurate results for small quenches, which
are defined as producing a small density of elementary excitations
through the quench \cite{CEF1,CEF2}. The leading terms are then
conjectured to be of the form
\bea
f_{0,1}&=&-i\tan\Delta_{k_0}\ln\left[ 
\frac{1+ie^{i\theta_{k_0}}\cos\Delta_{k_0}\tan\lambda}
{1+ie^{-i\theta_{k_0}}\cos\Delta_{k_0}\tan\lambda} \right]
\ ,\nn
f_{1,1}&=&\tan\Delta_{k_0} \left(i\ln\left[\frac{1+ie^{i\theta_{k_0}}\cos\Delta_{k_0}\tan\lambda}
{1+ie^{-i\theta_{k_0}}\cos\Delta_{k_0}\tan\lambda} \right]
-\frac{4\cos{\Delta_{k_0}}\tan{\lambda}\sin\theta_{k_0}}
{\sin^2\theta_{k_0}+\left( \cos\theta_{k_0}+i \cos\Delta_{k_0}\tan\lambda \right)^2}\right)
+{\cal O}(\sin^3(\Delta_{k_0})).
\label{fn1}
\eea
As we will see below, the contributions described by \fr{fn0} and
\fr{fn1} are sufficient to obtain an extremely accurate description of
$\chi^{(u)}(\lambda,\ell,t)$. The constant ${\cal C}$ can be fixed by
comparing the $t\to\infty$ limit of \fr{chiSP} to the result obtained
previously for the behaviour in the stationary state. For later
convenience we define two approximations as
\bea
\ln\chi^{(u)}_a(\lambda,\ell,t)=\ell \ln(\cos\lambda) + \frac{\ell}{2} 
\sum_{n=0}^2 \int_0^{2\pi}\frac{dk_0}{2\pi} \Theta(\ell-2n|v_k|t)
\left[1-\frac{2n|v_k|t}{\ell}\right]
\sum_{m=0}^a
\cos\big(2m\eps(k_0)t\big)f_{n,m}(\lambda,k_0)+{\cal C}\ ,
\label{chi12}
\eea
where $a=1,2$ and where we set $f_{2,1}=0$. The structure of the
integrand in our result \fr{chiSP} is reminiscent of that found in
connected two-point correlation functions \cite{CEF1} and entanglement
entropies \cite{FC08}. In the latter quantities it gives rise to a
``light-cone'' behaviour in developing connected correlations and the
spreading of entanglement respectively. In contrast to these cases the
expression \fr{chiSP} involves an infinite number of ``light-cone
structures'' with velocities that are integer multiples of the maximum
group velocity. Since the generating function involves complicated
sums over multi-point correlation functions on the interval $[1,\ell]$
this is not in contradiction with the celebrated Lieb-Robinson bound
\cite{lieb-1972}. The light-cone structure in connected two-point
correlators and entanglement entropies can be understood in terms of
simple semi-classical quasi-particle pictures \cite{cc-05,cc-06}.
It would be interesting to develop an analogous understanding for the
novel structure observed in the generating function, but this is
beyond the scope of the present paper.

%%%%%%%%%%%%%%%%%%%%%%%%%%%%%%%%%%%%%%%%%%%%%%%%%%%
\section{Accuracy of the asymptotic result}
\label{sec:ananum}
%%%%%%%%%%%%%%%%%%%%%%%%%%%%%%%%%%%%%%%%%%%%%%%%%%%
Our analytic result \fr{chiSP}, \fr{fn0}, \fr{fn1} gives the leading
contributions in the \emph{space-time scaling limit} \cite{CEF2}
$\ell,t\to\infty$, $\ell/t$ fixed. An important question is how good
this asymptotic result describes the behaviour of 
$\chi^{(u)}(\lambda,\ell,t)$ at small and intermediate times and
subsystem sizes. In order to answer this question we now turn
to a comparison between our analytical results \fr{chi12} and a direct
numerical evaluation of the determinant representation \fr{chiTI},
\fr{blockTI}. The numerical errors in the latter are negligible.
%%%%%%%%%%%%%%%%%%%%%%%%%%%%%%%%%%%%%%%%%%%%%%%%%%%%%%%
\subsection{Small-$\lambda$ regime}
\label{ssec:small}
\begin{figure}[ht!]
(a)\includegraphics[width=0.45\textwidth]{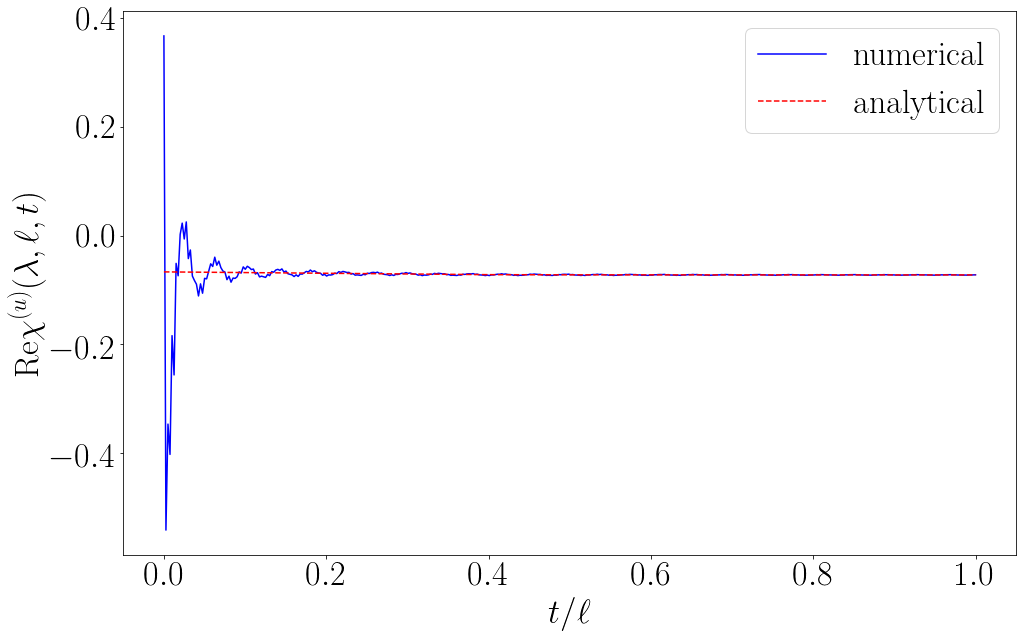}
\qquad
(b)\includegraphics[width=0.45\textwidth]{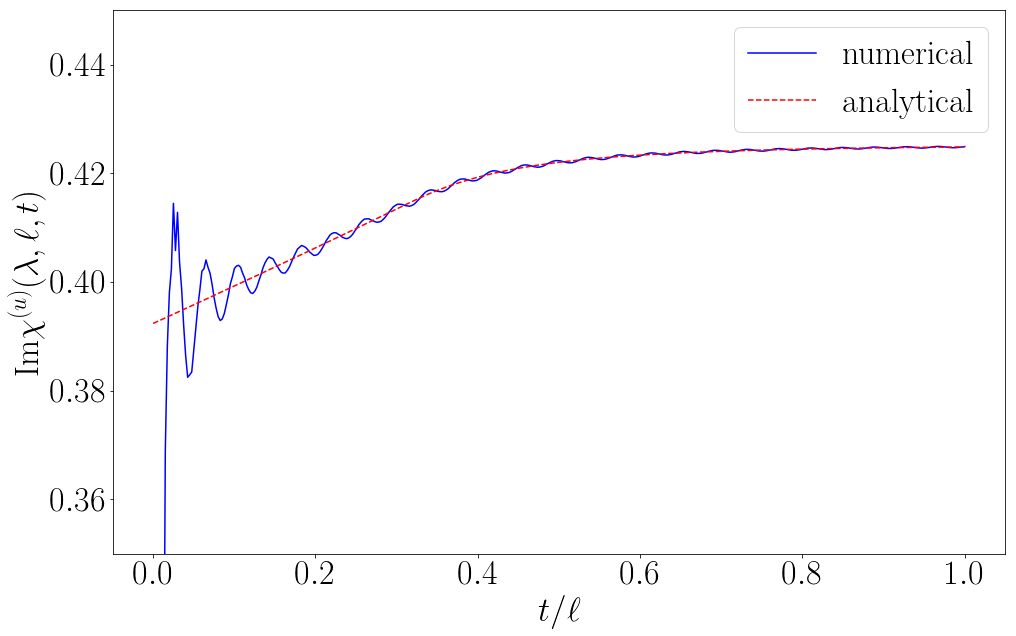}
\caption{Real and imaginary parts of the leading approximation
  $\chi^{(u)}_1(\lambda=0.1,\ell=200,t)$ for a transverse field quench
  quench from $h=0$ to $h=0.8$. The analytic approximation gives a
  good description only at late times.}
\label{fig:comp1}
\end{figure}

\begin{figure}[ht!]
(a)
\includegraphics[width=0.45\textwidth]{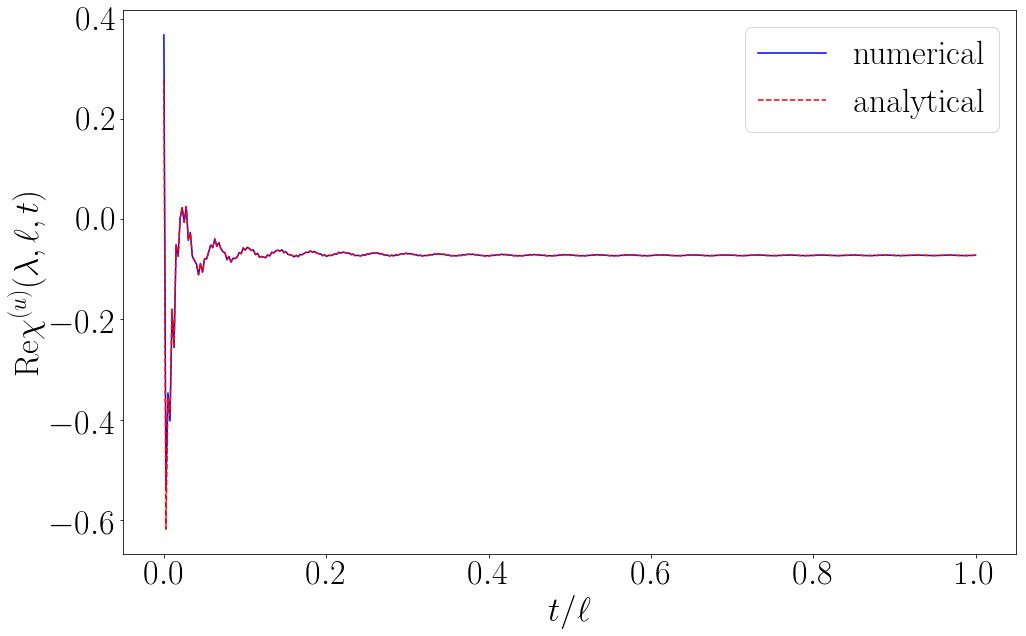}
\qquad
(b)
\includegraphics[width=0.45\textwidth]{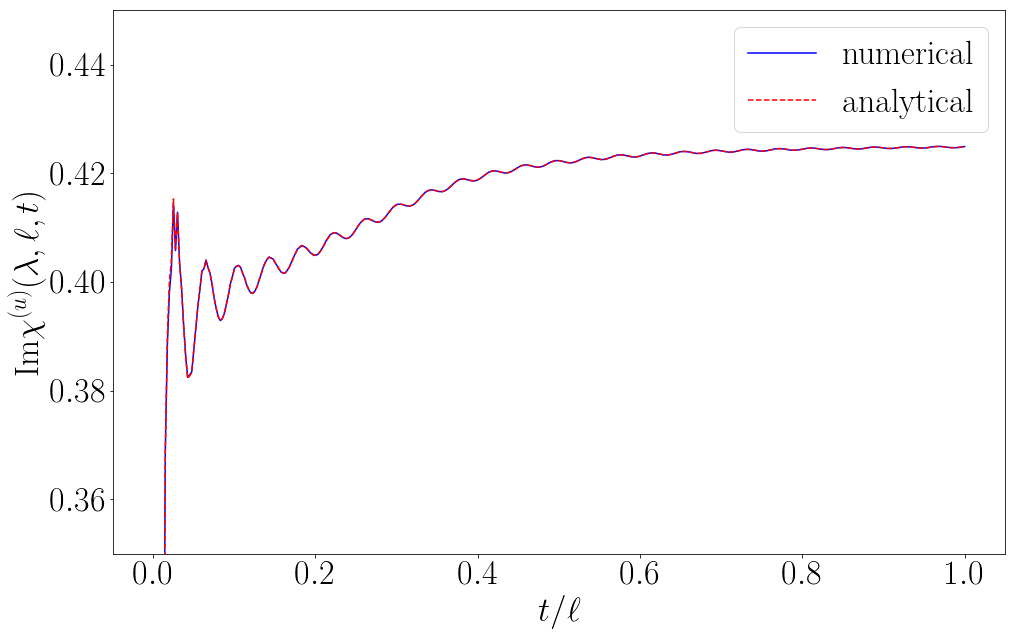}
\caption{Real and imaginary parts of $\chi^{(u)}_2(\lambda=0.1,\ell=200,t)$
for a transverse field quench quench from $h=0$ to $h=0.8$. The
analytical expression (red dashed line) is seen to be in excellent
agreement with the numerical results, which have negligible errors on
the scale of the figure.}
\label{fig:comp2}
\end{figure}

%%%%%%%%%%%%%%%%%%%%%%%%%%%%%%%%%%%%%%%%%%%%%%%%%%%%%%%
A representative comparison between the analytical results
$\chi^{(u)}_{1,2}(\lambda,\ell,t)$ for small values of $\lambda$ and
numerics is shown in Fig~\ref{fig:comp1} and \ref{fig:comp2}.
We see that $\chi^{(u)}_1(\lambda,\ell,t)$ reproduces the numerics
very well at late times after the quench. In contrast, the oscillatory
behaviour at short times is clearly not captured.
The improved approximation $\chi^{(u)}_2(\lambda,\ell,t)$ \fr{chi12}
is seen to be in excellent agreement with the numerics.

By construction the oscillatory part of the analytic result is most accurate over the entire range
of the ``counting parameter'' $\lambda$ when $\sin\Delta_{k_0}$ is
small, i.e. for small quenches. For quenches where $\sin\Delta_{k_0}$
is no longer small we still find excellent agreement between the 
analytic and numerical results as long as $\tan(\lambda)$ is
small. This can be understood by noting that for such values of $\lambda$ the
infinite sum in \eqref{chipowerseries} is dominated by the first few
terms, i.e. small values of $n$. On the other hand, higher orders of
$\sin\Delta_{k_0}$ only emerge for larger values of $n$. Therefore the
leading order result in $\sin\Delta_{k_0}$ already provides a very
good approximation in the small-$\tan(\lambda)$ regime even when
$\sin\Delta_{k_0}$ is not small. This observation is of significant
practical importance:
\begin{figure}[ht!]
(a)\includegraphics[width=0.45\textwidth]{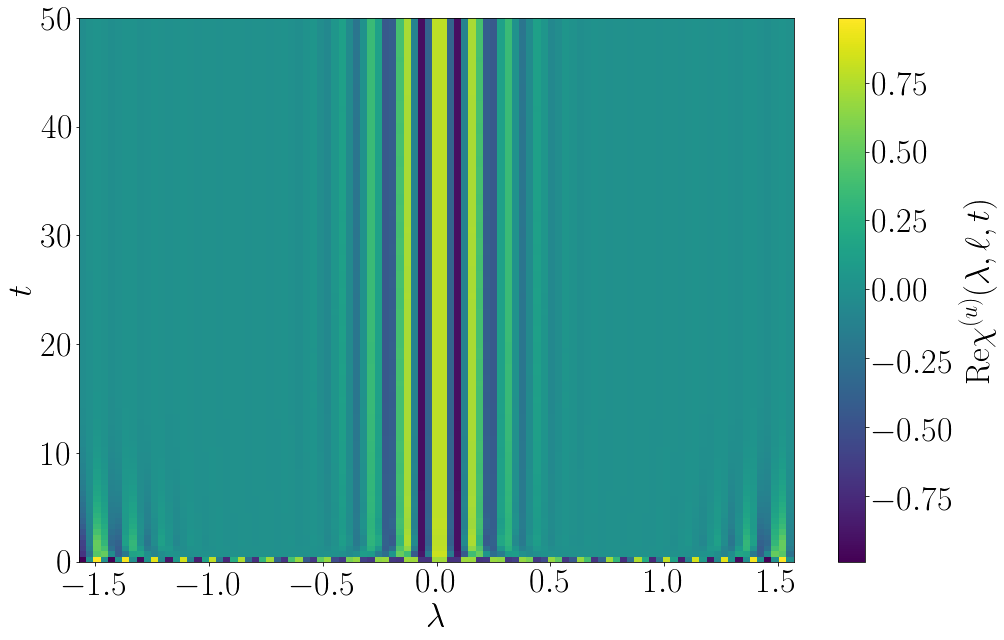}
\qquad
(b)\includegraphics[width=0.45\textwidth]{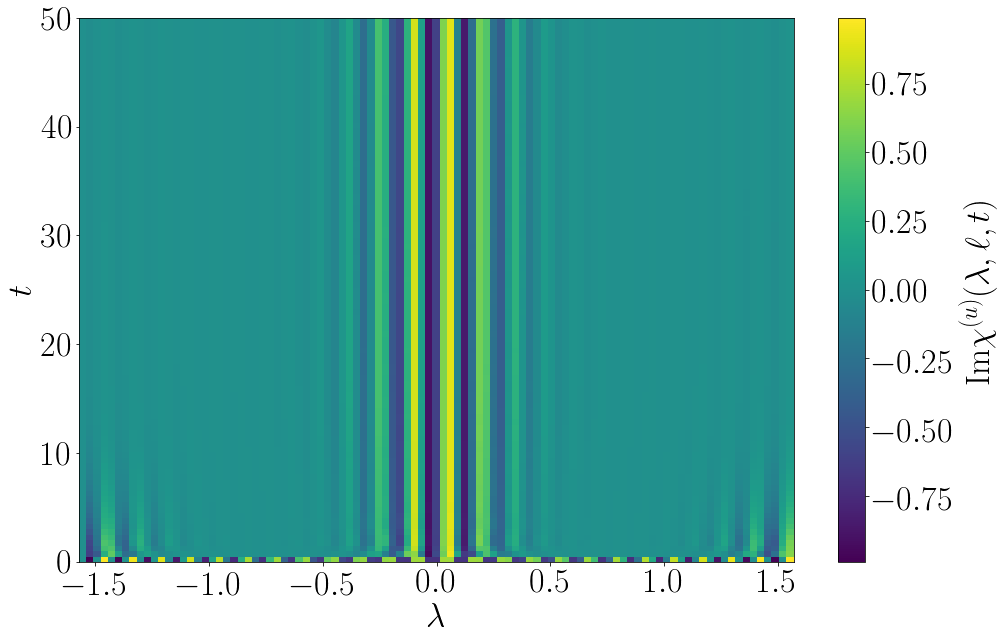}
\caption{(a) Real and (b) imaginary parts of $\chi^{(u)}(\lambda,\ell=50,t)$ 
as functions of $\lambda$ and $t$ for a transverse field quench
from $h=5$ to $h=1.5$. We observe that the characteristic function is
small unless $\lambda$ is small. The behaviour for quenches within the
ferromagnetic phase and quenches between the phases is similar.}
\label{Fig:FCS_lambda_t}
\end{figure}
As shown in Fig.~\ref{Fig:FCS_lambda_t} in a particular example $|{\rm
  Re}\chi^{(u)}(\lambda,\ell,t)|$ and $|{\rm Im}\chi^{(u)}(\lambda,\ell,t)|$ are
largest in the vicinity of $\lambda=0$ (except at short times). This
implies that the corresponding probability distribution, which is the
object we are ultimately interested in, will be dominated by the
small-$\lambda$ regime. As a consequence \fr{chiSP}, \fr{fn0},
\fr{fn1} provide a good approximation for the calculation of
$P^{(u)}_w(m)$ for all quenches. 
%%%%%%%%%%%%%%%%%%%%%%%%%%%%%%%%%%%%%%%%%%%%%%%%%%%%%%%
\subsection{Large-$\lambda$ regime}
\label{ssec:large}
\begin{figure}[ht!]
  (a)\includegraphics[width=0.45\textwidth]{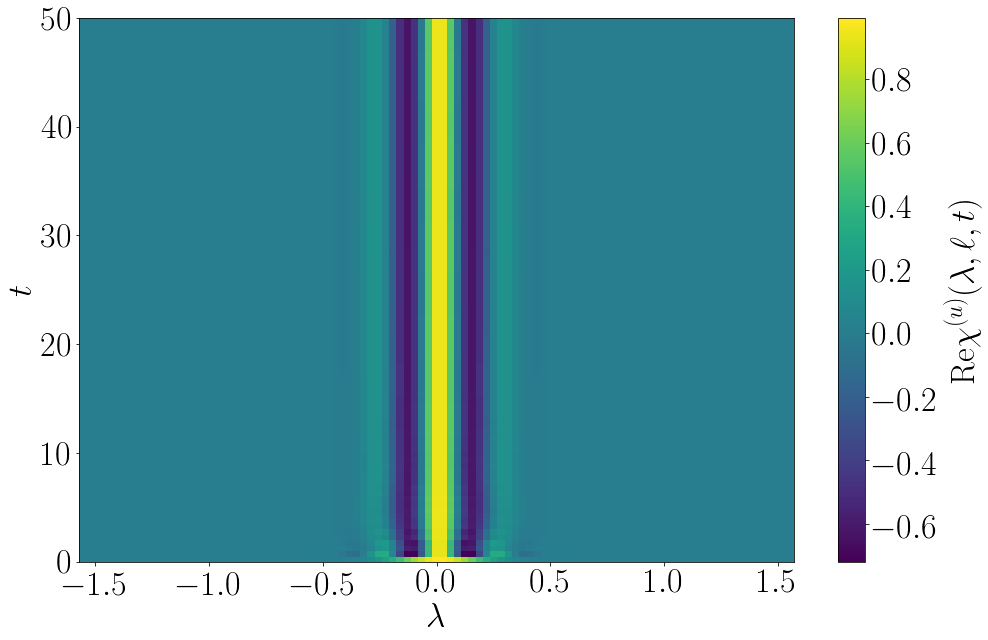}
  \qquad
  (b)\includegraphics[width=0.45\textwidth]{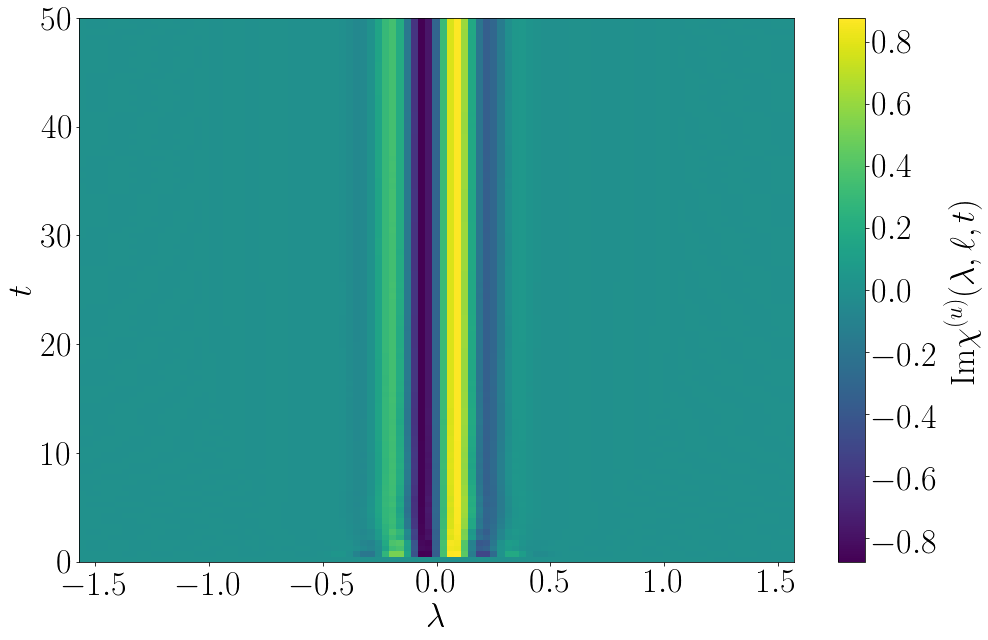}
  \caption{Real (a) and imaginary (b) parts of
    $\chi^{(u)}(\lambda,\ell=50,t)$ for a quench within the
    ferromagnetic phase from $h_0=0.2$ to $h=0.8$. The dominant
    contribution to the probability distribution arises from the
    small-$\lambda$ regime.}
  \label{Fig:chi_lambda_t_ferro}
\end{figure}

%%%%%%%%%%%%%%%%%%%%%%%%%%%%%%%%%%%%%%%%%%%%%%%%%%%%%%%
In the large-$\lambda$ regime we have to distinguish between the cases
where the symbol in the stationary state has zero or non-zero winding
number, \emph{c.f.} section \ref{sec:stationarystate}.
The first case covers quenches to the paramagnetic phase. Here we find
that our analytic result is again in good agreement with numerics.
The second scenario applies to quenches to the ferromagnetic phase and
$\lambda>\lambda_c(h_0,h)$. We have shown in section \ref{sec:scaling}
that there is no good scaling collapse in this regime of counting
parameters for the moderate subsystem sizes and times of interest
here. It should therefore not come as a surprise that the asymptotic
result does not provide a good approximation in this regime. Presumably
\fr{chiSP}, \fr{fn0}, \fr{fn1} no longer hold in this regime because
the analytic continuation of the power series expansion of the
logarithm \eqref{chipowerseries} becomes non-trivial in this case.
In practice the failure of the analytic approach to give a good
account of the generating function in this parameter regime is
irrelevant as $\chi^{(u)}(\lambda,\ell,t)$ itself is extremely small
and makes a negligible contribution to the probability distribution.
As shown in Fig.~\ref{Fig:chi_lambda_t_ferro}, the main contribution
to the latter, which after all is our object of interest, arises from
the small-$\lambda$ regime of the generating function, which is well
approximated by our analytic expressions. 

%%%%%%%%%%%%%%%%%%%%%%%%%%%%%
\subsection{Relative errors}
\label{ssec:errors}
%%%%%%%%%%%%%%%%%%%%%%%%%%%%
In order to provide a more quantitative discussion of the quality
of the approximate results \fr{chi12} we consider the relative
errors 
\be
r_{1,2}(\lambda,\ell,t)=
\left|1-\frac{\ln\big(\chi^{(u)}_{1,2}(\lambda,\ell,t)\big)}
{\ln\big(\chi^{(u)}_{\rm num}(\lambda,\ell,t)\big)}\right|\ ,
\ee
where $\chi^{(u)}_{\rm 1,2/num}(\lambda,\ell,t)$ are respectively the
analytic approximations \fr{chi12} and the result of the numerical
computation of the determinant representation \fr{chiTI}, \fr{blockTI}. 
\begin{figure}[ht!]
(a)\includegraphics[width=0.45\textwidth]{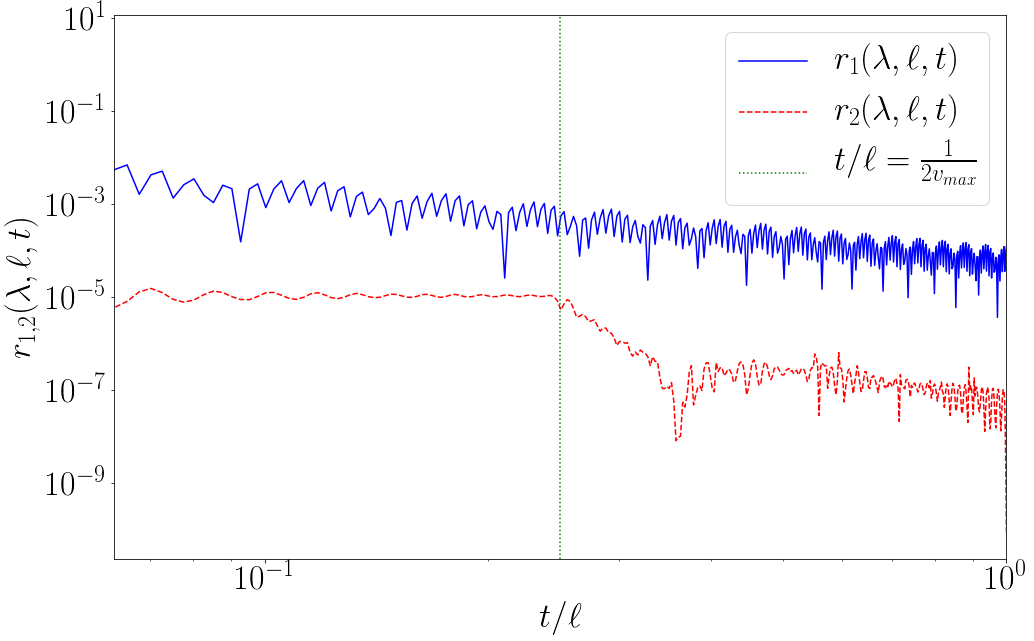}
\qquad
(b)\includegraphics[width=0.45\textwidth]{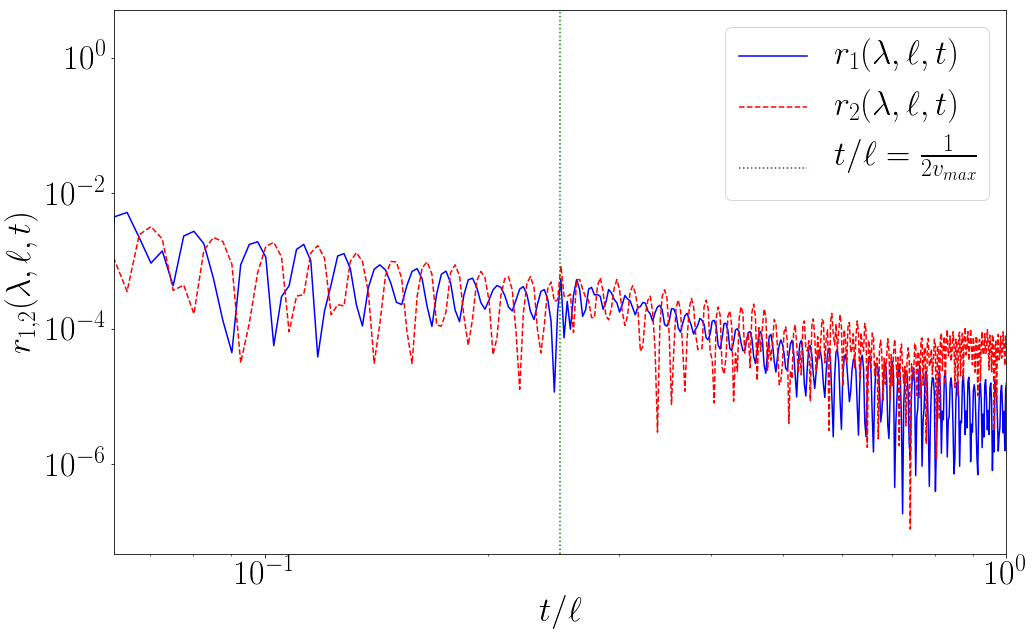}
\caption{(a) Relative errors $r_{1,2}(\lambda=0.1,\ell=200,t)$ for a quench
within the paramagentic phase from $h_0=5$ to $h=1.5$.
(b) same for $r_{1,2}(\lambda=1.4,\ell=200,t)$.}
\label{fig:acidtest}
\end{figure}
\begin{figure}[ht!]
(a)\includegraphics[width=0.45\textwidth]{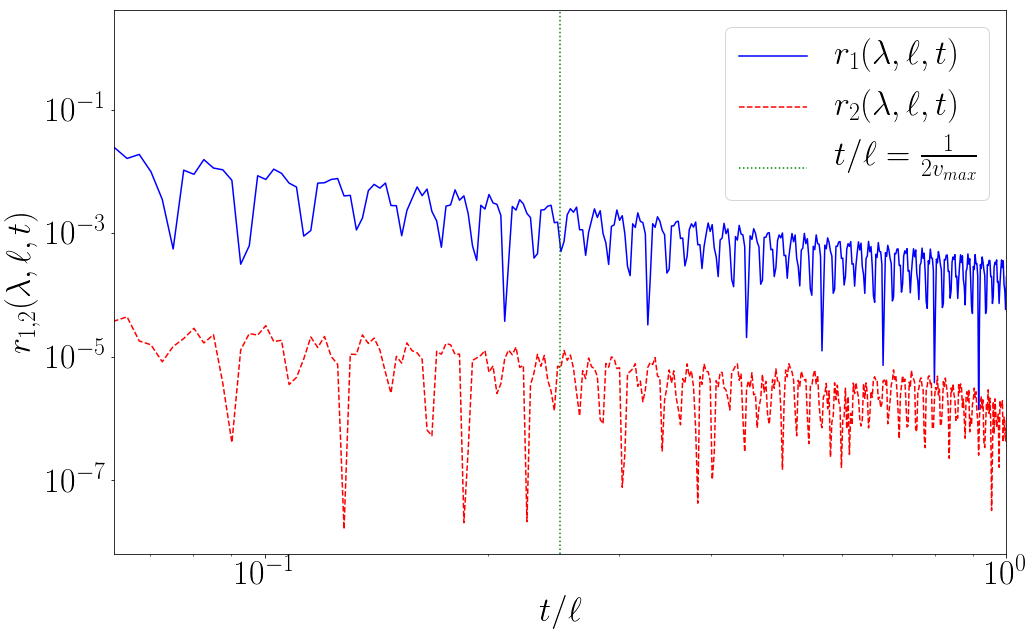}
\qquad
(b)\includegraphics[width=0.45\textwidth]{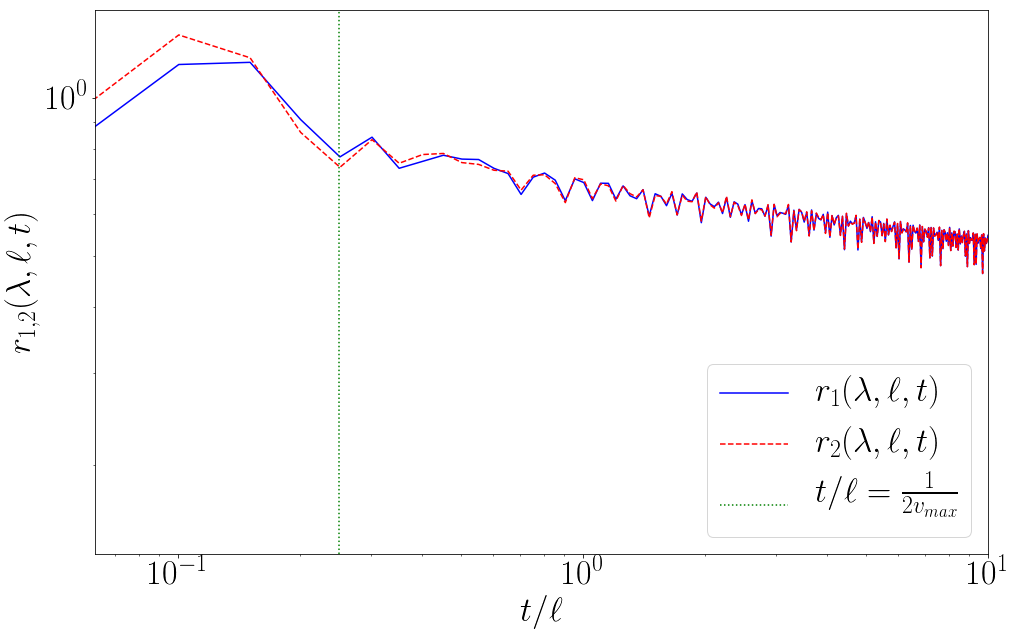}
\caption{(a) Relative errors $r_{1,2}(\lambda=0.1,\ell=200,t)$ for a quench
within the ferromagnetic phase from $h_0=0$ to $h=0.8$.
(b) same for $r_{1,2}(\lambda=1.4,\ell=200,t)$.}
\label{fig:acidtest_ferro}
\end{figure}
In Fig.~\ref{fig:acidtest} we plot the time dependence of the relative
errors for a quench from $h_0=5$ to $h=1.5$ for a subsystem of size
$\ell=200$ and two values of the counting parameter $\lambda$.
The maximal value of $\sin(\Delta_{k_0})$ within the domain of
integration  approximately $0.54$, which means that higher orders in
$f_{1,1}$ can be important. As we have argued above, this will be the
case if $\tan(\lambda)$ is not small. In Fig.~\ref{fig:acidtest} (a)
$\lambda=0.1$ is taken to be small, and the quality of both
approximations $\chi^{(u)}_{1,2}(\lambda,\ell,t)$ is seen to be excellent. In 
Fig.~\ref{fig:acidtest} (b) the counting parameter $\lambda=1.4$ is
taken to be large. This leads to a significantly larger error, which
is however still fairly small and also decays in time. We see that the
analytic results provide a good approximation for all values of $\lambda$.

We now turn to a parameter regime, in which our analytic results no
longer provide a uniformly good approximation for all values of the
counting parameter $\lambda$. Fig.~\ref{fig:acidtest_ferro} shows
results for a quench from $h_0=0.2$ to $h=0.8$. The maximal value of 
$\sin\Delta_{k_0}$ in the integration range is now $0.71$ so that
higher orders in $f_{1,1}$ can again be important. For small values of
$\lambda$ the relative errors of both analytical approximations
are small and decreasing in time. On the other hand
$\chi^{(u)}_{1,2}(\lambda,\ell,t)$ cease to provide accurate
approximations for large values of $\lambda$ with
$\lambda>\lambda_c(h_0,h)$ as can be seen in
Fig.~\ref{fig:acidtest_ferro} (b). However, we want to stress once
more that $\chi^{(u)}(\lambda,200,t)$ itself is extremely small in
this parameter regime and makes only a negligible contribution to the
probability distribution.

%%%%%%%%%%%%%%%%%%%%%%%%%%%%%%%%%%%%%%
\subsection{Probability distributions}
\label{ssec:probs}
\begin{figure}[ht!]
(a)\includegraphics[width=0.45\textwidth]{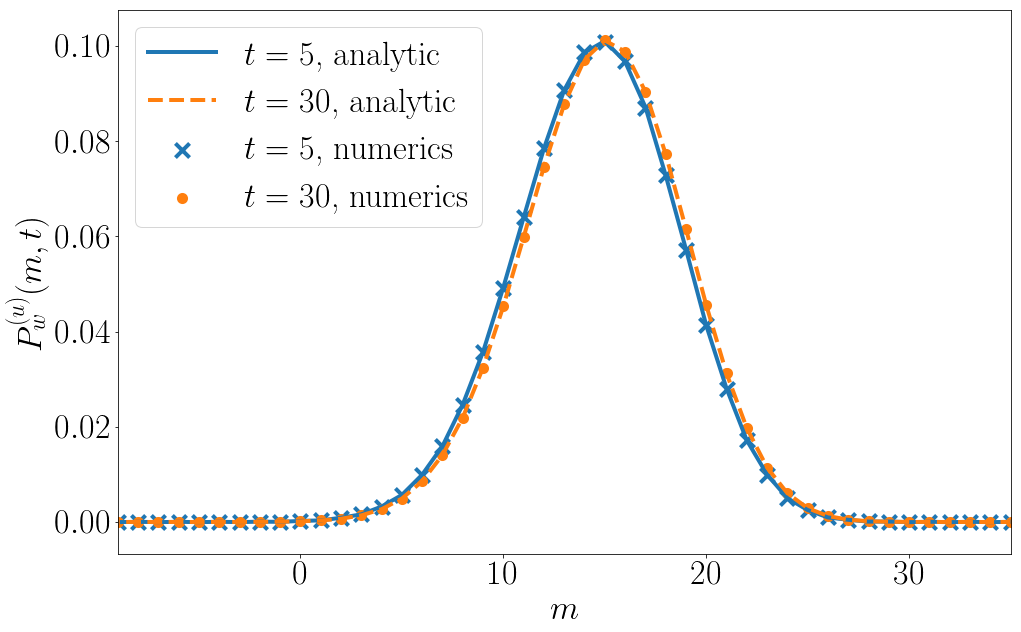}
\qquad
(b)\includegraphics[width=0.45\textwidth]{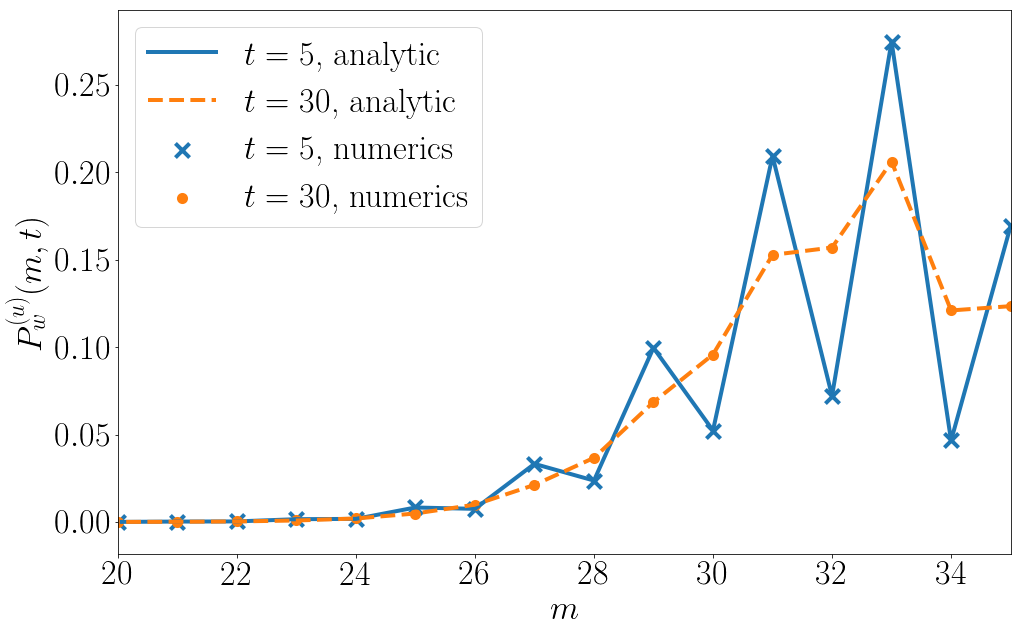}
\caption{Comparison of the asymptotic expression for $P_w^{(u)}(m,t)$
obtained from eqns \fr{chiSP}, \fr{fn0}, \fr{fn1} (solid lines) to
numerics (symbols) for transverse field quenches with (a)
$h_0=0.2$ and $h=0.8$ and (b) $h_0=5$ and $h=2$. The agreement is seen
to be excellent.}
  \label{Fig:prob_ana_num}
\end{figure}

%%%%%%%%%%%%%%%%%%%%%%%
An asymptotic expansion for the probability distribution
$P_w^{(u)}(m,t)$ can be obtained by Fourier transforming the
generating function, \emph{cf.} Eq. \fr{Pweights}. As expected on the
basis of the discussion above, we find that the analytic result
becomes very accurate at sufficiently late times for \emph{all}
quenches. At intermediate and short times we still find excellent
agreement between the analytical and numerical results for quenches
originating the ferromagnetic, see e.g. Fig.~\ref{Fig:prob_ana_num}
(a). 
For quenches from the paramagnetic phase the analytic result is
an excellent agreement with numerics at short and intermediate times
as long as the quench is ``small''. In practice this covers all
quenches within the paramagnetic phase as long as $h$ is not very
close to $1$. For other quenches the corrections to the $f_{1,1}$ term
in \fr{fn1} will become significant at short and intermediate times.

%%%%%%%%%%%%%%%%%%%%%%%%%%%%%
\section{Conclusions}
\label{sec:summ}
%%%%%%%%%%%%%%%%%%%%%%%%%%%%%
We have analysed the full counting statistics of the transverse and staggered 
magnetization of a subsystem in the thermodynamic limit of the
transverse field Ising chain. We derived a convenient determinant
representation for the corresponding generating functions
$\chi^{(u,s)}(\lambda,\ell,t)$. We first considered the FCS in
equilibrium states and showed that the probability distributions are
always non-Gaussian except in the limit of infinite subsystem size at
finite temperature. We determined the temperature and field dependence of 
the generating function as well as the first few cumulants. We then
moved on to the main focus of our work, the calculation of the FCS after
quantum quenches. We considered two quench protocols: transverse field
quenches and evolution starting from a classical N\'eel state. We
first determined the FCS in the stationary states reached at late
times. The probability distributions are again non Gaussian, except in
the limit of infinite subsystem size.  
We analyzed the time evolution of the probability distributions
$P^{(u,s)}(m,t)$ for a variety of quenches by numerically evaluating
the exact determinant representation for the generating function (the
numerical errors incurred are negligible). For transverse field
quenches originating in the paramagnetic phase $P^{(u,s)}(m,t)$ showed
interesting smoothing and broadening behaviour in time. In contrast,
$P^{(u,s)}(m,t)$ displayed a simpler behaviour 
for quenches originating in the ferromagnetic phase.
In the case of a N\'eel quench $P^{(s)}(m,t)$ encoded detailed
information on the restoration of translational invariance. The
numerical approach provided us with evidence for the existence of a
\emph{scaling regime} for the generating function in which we observed
data collapse according to the scaling form \eqref{scaling}. This is
turn allowed us to proceed with the derivation of the main result of
our work: the analytic expression \eqref{chiSP} for the FCS after
transverse field quenches in the \emph{space-time scaling limit}
$t,\ell\to\infty$, $t/\ell$ fixed. This was achieved by a substantial
generalization of the multi-dimensional stationary phase approximation
method of Refs~\cite{FC08,CEF2}. We performed a careful comparison of
our analytic results to numerics (that has negligible errors) and
found excellent agreement on the level of the probability
distributions for all cases considered. We observed that the expression for
the generating function exhibits an interesting multiple light-cone
structure that has no analog in either correlation functions of local
observables \cite{CEF1} or in the entanglement entropy \cite{FC08}. An
interesting open question is whether this structure can be understood
in terms of the kind of semiclassical quasi-particle picture that has
been successfully employed to explain the main features observed in
the dynamics of both entanglement \cite{cc-05} and correlations \cite{cc-06}. 

Our work provides the first analytic results for FCS after quantum
quenches and hopefully will pave the way for further studies. Here we
have focussed on the FCS for the transverse magnetisation. It would be
very interesting to determine the FCS for the longitudinal
magnetisation, which is the order parameter characterising the Ising quantum phase
transition. A more straightforward but interesting extension would be
to study certain observables in free fermion models with long-range hopping
and/or pairing \cite{regemortel15,buyskikh16,lepori17}. Similarly, the
probability distribution of the (smooth) subsystem magnetisation in
the spin-1/2 Heisenberg XXZ chain should be calculable both at finite
temperatures \cite{Gohmann13} and in the stationary states after
certain quantum quenches
\cite{CE_PRL13,GGE_XXZ_Amst,QANeel,XXZung,XXZunglong,GGE_int,IQNB,pvc-16,pvcr-16}.  
For quantum quenches in the regime where bosonization provides a
good approximation \cite{cce-15} the full time evolution of the
probability distribution for certain observables can be obtained in a
straightforward way. Finally, the case of integrable chains of higher spin
could be studied both in equilibrium \cite{jls-89,b-83} and after a
quench \cite{pvc-16,mbpc-17}.

%%%%%%%%%%%%%%%%%%%%%%%%%%%%%
\acknowledgements
%%%%%%%%%%%%%%%%%%%%%%%%%%%%%
This work was supported by the EPSRC under grant EP/N01930X (FHLE) and
by the Clarendon Scholarship fund (SG). 
We are grateful to the Erwin Schr\"odinger International Institute for
Mathematics and Physics for hospitality and support during the
programme on \emph{Quantum Paths}. 

\appendix
%%%%%%%%%%%%%%%%%%%%%%%%%%%%%%%%%%%%%%%%%%%%%
\section{Asymptotics of block Toeplitz matrices}
\label{app:Szego}
%%%%%%%%%%%%%%%%%%%%%%%%%%%%%%%%%%%%%%%%%%%%%%

Let $T_\ell$ be a general block Toeplitz matrix with elements $(T_\ell)_{ln}=t_{l-n}$.
The \emph{symbol} $\tau(e^{ik})$ of $T_\ell$ is defined by 
\begin{align}
t_n \equiv \int_{0}^{2\pi} \frac{dk}{2\pi} \tau(e^{ik}) e^{-ink}.
\end{align}
In cases where the symbol has winding number zero, the large-$\ell$
asymptotics of the determinant of $T_\ell$ is (under certain
conditions) given by \cite{widom} 
\begin{align}
  \ln\det {T_\ell}  = \ell \int_{0}^{2\pi} \frac{dk}{2\pi}
  \ln{\det{\tau(e^{ik})}} + \det{T(\tau^{-1})T(\tau)} + o(1).
\end{align}
Here $T(\tau)$ denotes an infinite Toeplitz matrix with symbol $\tau$.
In the case where the block-size is $1$, this reduces to the
Szeg{\H{o}} limit theorem 
\begin{align}
\ln\det {T_\ell}  = \ell \int_{0}^{2\pi} \frac{dk}{2\pi} \ln{\tau(e^{ik})} +\sum_{q\geq 1} q \;(\ln \tau)_q(\ln \tau)_{-q}+ o(1),
\end{align}
where 
\begin{align}
  (\ln \tau)_q = \int_{0}^{2\pi} \frac{dk}{2\pi} \ln \tau(e^{ik})
  e^{-ikq}\ .
\end{align}
The large $\ell$ asymptotics of Toeplitz determinants in cases where
the symbol $\tau$ has winding number $\pm 1$ is given by
\cite{boettcherwidom,CEF2}  
\begin{align}
  \ln\det {T_\ell}  = \ell \int_{0}^{2\pi} \frac{dk}{2\pi}
  \ln{a(e^{ik})} +\sum_{q\geq 1} q \;(\ln a)_q(\ln a)_{-q}+ \ln
  \int_{0}^{2\pi} \frac{\id k}{2\pi} e^{-i\ell k}
  \frac{a_{-}(e^{ik})}{a_{+}(e^{ik})} + o(1)\ ,
\end{align}
where 
\begin{align}
  a(e^{ik}) \equiv -e^{\mp ik} \tau(e^{ik})= \exp{\sum_{j=1}^\infty
    (\ln a)_{\pm j} e^{\pm ijk}}\ .
\end{align}

%%%%%%%%%%%%%%%%%%%%%%%%%%%%%%%%%%%%%%%%%%%%%%%%%%%%%%
\section{Perturbation theory around the $h\to\infty$ limit}
\label{appendix:perturb}
%%%%%%%%%%%%%%%%%%%%%%%%%%%%%%%%%%%%%%%%%%%%%%%%%%%%%%
We have seen that the probability distributions $P^{(u,s)}(m,t)$
exhibit an even/odd structure in $m$ for short times after quenches
starting in the paramagnetic phase. In this appendix we show that this
structure can be understood in perturbation theory around the
$h\to\infty$ limit. For simplicity we consider the probability
distribution $P^{(u)}(m)$ in the ground state at $h\gg 1$. In the limit
$h\to\infty$ the ground state is the saturated ferromagnetic state
along the transverse field direction 
\be
|0\rangle^{(0)}=|\uparrow\dots\uparrow\rangle\ .
\ee
Hence
\be
{}^{(0)}\langle 0|e^{i\lambda S^z_u(\ell)}|0\rangle^{(0)}=e^{i\lambda\ell}\ .
\ee
The corresponding probability distribution is a delta function at
$m=-\ell/2$. The other eigenstates of $\sum_j\sigma^z_j$ are denoted
by $|n\rangle^{(0)}$. The leading correction to the generating
function arises at second order in perturbation theory in
$H_1=\sum_j\sigma^x_j\sigma^x_{j+1}$. The relevant corrections to the
ground state are 
\be
|0\rangle^{(2)}=|0\rangle^{(0)}+\sum_{n\neq 0}
|n\rangle^{(0)}
\frac{{}^{(0)}\langle n|H_1|0\rangle^{(0)}}{E^{(0)}_0-E_n^{(0)}}
-\frac{1}{2}|0\rangle^{(0)}\sum_{n\neq 0}
\frac{\big|{}^{(0)}\langle n|H_1|0\rangle^{(0)}\big|^2}{(E^{(0)}_n-E^{(0)}_0)^2}
+\dots
\ee
Substituting this into the expression for the generating function
gives
\bea
{}^{(2)}\langle 0|e^{i\lambda
  S^z_u(\ell)}|0\rangle^{(2)}&=&e^{i\lambda\ell}
\left[1-\sum_{n\neq 0}
\frac{\big|{}^{(0)}\langle
  n|H_1|0\rangle^{(0)}\big|^2}{(E^{(0)}_n-E^{(0)}_0)^2}\right]
+\sum_{n\neq 0}{}^{(0)}\langle n|e^{i\lambda S^z_u(\ell)}|n\rangle^{(0)}
\Bigg|\frac{{}^{(0)}\langle n|H_1|0\rangle^{(0)}}{E^{(0)}_0-E_n^{(0)}}\Bigg|^2.
\eea
In order for ${}^{(0)}\langle n|H_1|0\rangle^{(0)}$ to be non-zero the
product state $|n\rangle^{(0)}$ must have precisely two overturned
spins. Let us denote their positions by $j$ and $j+1$. For $\ell\geq 2$ we
then have 
\be
{}^{(0)}\langle n|e^{i\lambda
  S^z_u(\ell)}|n\rangle^{(0)}=\begin{cases}
e^{i\lambda(\ell-4)} & \text{if } 1\leq j<\ell\\
e^{i\lambda(\ell-2)} & \text{if } j=0\text{ or }\ell\\
e^{i\lambda\ell} & \text{else}. \\
\end{cases}
\ee
This gives
\bea
{}^{(2)}\langle 0|e^{i\lambda
  S^z_u(\ell)}|0\rangle^{(2)}&=&e^{i\lambda\ell}
\left[1-\frac{\ell+1}{16h^2}\right]
+\frac{2}{16h^2}e^{i\lambda(\ell-2)}
+\frac{\ell-1}{16h^2}e^{i\lambda(\ell-4)}\ .
\eea
The corresponding probability distribution is
\be
P^{(u)}(m)\Bigg|_{\rm PT}=\left[1-\frac{\ell+1}{16h^2}\right]\delta(m-\ell/2)
+\frac{2}{16h^2}\delta(m+1-\ell/2)+\frac{\ell-1}{16h^2}\delta(m+2-\ell/2).
\ee
This is seen to exhibit an even/odd effect as the corrections for
$m=\ell/2$ mod 2 are proportional to the subsystem size.

\end{document}